\documentclass[final,5p,twocolumn]{elsarticle}

\usepackage{hyperref}
\usepackage{graphicx}
\journal{BBA-Biomembranes}

 \biboptions{sort&compress}
\newcommand{\dir}{./}
\newcommand{\etal}{et al.}
\newcommand{\ud}{{\rm d}}

\begin{document}

\begin{frontmatter}

\title{Physical mechanisms of micro- and nanodomain 
     formation in multicomponent lipid membranes}
%\title{Elsevier \LaTeX\ template\tnoteref{mytitlenote}}

\author{Friederike Schmid}
\address{Institute of Physics, Johannes Gutenberg University, 55099 Mainz}




\begin{abstract}

This article summarizes a variety of physical mechanisms proposed in the
literature, which can generate micro- and nanodomains in multicomponent lipid
bilayers and biomembranes. It mainly focusses on lipid-driven mechanisms that
do not involve direct protein-protein interactions. Specifically, it considers
(i) equilibrium mechanisms based on lipid-lipid phase separation such as
critical cluster formation close to critical points, and multiple domain
formation in curved geometries, (ii) equilibrium mechanisms that stabilize
two-dimensional microemulsions, such as the effect of linactants and the effect
of curvature-composition coupling in bilayers and monolayers, and (iii)
non-equilibrium mechanisms induced by the interaction of a biomembrane with the
cellular environment, such as membrane recycling and the pinning effects of the
cytoplasm. Theoretical predictions are discussed together with simulations and
experiments.  The presentation is guided by the theory of phase transitions and
critical phenomena, and the appendix summarizes the mathematical background in
a concise way within the framework of the Ginzburg-Landau theory.

\end{abstract}

\begin{keyword}
Membrane \sep
Cholesterol \sep
Lipid domains \sep
Lipid phase separation \sep
Lipid sorting \sep
Curvature \sep
Two-dimensional microemulsions\sep
Membrane recycling\sep
Cytoplasm
%\texttt{elsarticle.cls}\sep \LaTeX\sep Elsevier \sep template
%\MSC[2010] 00-01\sep  99-00
\end{keyword}

\end{frontmatter}

\section{Introduction}
\label{sec:introduction}

Ever since Simons and Ikonen first coined the term ''rafts'' to describe
certain lateral inhomogeneities in lipid membranes \cite{SI97}, the question
whether rafts exist in vivo and why they form has been the subject of a lively
and often controversial debate in the biophysics community \cite{BL98, BL98b,
BL00, JD99, Edidin03, Munro03, Pike03, Shaw06, Hancock06, JMA07, LS10,
Leslie11, KSK11, HF11, Kraft13, SBF15}. Even the meaning of the word ''raft''
has long remained vague, until in 2006 the participants of a Keystone Symposium
on Lipid Rafts and Cell function have formulated a ''consensus definition''
\cite{Pike06}: {\em ''Membrane rafts are small (10-200 nm), heterogeneous,
highly dynamic, sterol- and sphingolipid-enriched domains that compartmentalize
cellular processes. Small rafts can sometimes be stabilized to form larger
platforms through protein-protein and protein-lipid interactions.''} This
description now serves as a reference for the identification  of raft-like
structures, even though not all structures that have been associated with rafts
fulfill all criteria. For example, ''rafts'' do not always contain
sphingolipids \cite{LCC10,LPC13}. 

The raft concept is supported by increasing experimental evidence -- e.g, from
tracking of lipid diffusion \cite{ERM09}, or from superresolution microscopy
\cite{MAD11,OMW12,DDK15} -- that lipid membranes are heterogeneous on a
nanometer scale \cite{VSM03}. Nano- and microdomains in membranes have been
observed in a large range of organisms, including prokaryotic cells
\cite{LCC10,LPC13}, single-cell organisms \cite{TP13}, and plant cells
\cite{JKS14}. Motivated by the observation that sphingolipids (sphingomyelins,
glycosphingolipids, ceramides) -- which participate in cellular signalling --
tend to enrich in raft domains \cite{ZLB09,WeS09,GA09}, it has been speculated
that rafts may serve a biological purpose in cell-cell recognition and signal
transduction \cite{HH07,ZLB09}. On the other hand, raft-like structures were
also observed in prokaryotic membranes which do not contain sphingolipids
\cite{LCC10}. 
 
One class of lipids which seems to be very prominently involved in raft
formation is the sterol class \cite{NHV09,RM13}. With few exceptions
\cite{FKL13}, higher sterols such as cholesterol and ergosterol are typically
enriched in rafts. Recently, LaRocca \etal\ performed a systematic substitution
study on prokaryotic membranes, and reported that domain formation was
suppressed if cholesterol was depleted or substituted with the wrong sterol
\cite{LPC13}. Since the cholesterol molecule with its rigid structure has an
ordering effect on the acyl chains of the surrounding lipids
\cite{Leathe25,Mouritsen05,RPV09}, the apparently dominant role of cholesterol
suggests an interpretation of rafts in terms of a local nucleation of ordered
cholesterol-rich ''liquid ordered'' ({\em lo}) domains in a ''liquid
disordered'' ({\em ld}) sea of cholesterol-poor phase \cite{RM13}.  ''Inverse
domains'' have also been observed \cite{WaS09}, where highly disordered domains
with a high content of polyunsaturated fatty acids segregate from a
cholesterol-rich environment. 

The idea that lipid-cholesterol bilayers may demix into ''liquid ordered'' and
''liquid disordered'' states had been put forward already in 1987 by Ipsen
\etal \cite{IKM87} based on experimental data by Vist \etal \cite{VD90} (see
Sec.\ \ref{sec:phase_separation}).  In contrast to equilibrium phase
separated domains, however, lipid rafts are small and transient structures. The
question why such domains should form has intrigued scientists for some time.
In vivo, membranes are filled with proteins and lipid domains are typically
correlated with protein clusters.  Therefore, it has been argued that the
observed membrane heterogeneities could be driven by protein-protein
interactions alone \cite{Destainville10}.  If ''raft proteins'' associate with
''raft lipids'' such as cholesterol, it seems conceivable that a protein
cluster could nucleate a liquid ordered {\em lo} lipid domain in its vicinity.
On the other hand, recent experiments by Sevcsik \etal \cite{SBF15} have shown
that immobilized lipid anchored raft proteins do not seem to have a measurable
effect on the membrane environment.  This suggests that the formation of lipid
domains is primarily driven by the lipids and not by the membrane proteins.

Indeed, nanostructures and microstructures are also observed in pure model
lipid bilayers. One prominent example in one-component bilayers is the
modulated ''ripple phase'' $P_\beta'$, which generically emerges in the
transition region between the fluid phase $L_\alpha$ and a tilted gel phase
$L_\beta'$ \cite{KC98, KC02, KTL00, LKC02, LS07, SDL14}  (see Sec.\
\ref{sec:phase_separation}). Here and throughout, the term ''modulated'' refers
to periodic or quasi-periodic patterns -- in this case striped patterns with
periodicities of the order of 10 nm. In multicomponent lipid bilayers,
experimental evidence for the existence of nanoscopic cholesterol-rich domains
has been provided by F\"orster resonance electron transfer (FRET) and electron
spin resonance (ESR) experiments \cite{HWG10, PHD13}, neutron scattering
methods \cite{HPP13, AMD13, RM13, TMA14, NCM15}, interferometric scattering
microscopy \cite{DDK15}, and atomic force microscopy \cite{CH13, KhHP15}.
Micron-size domain patterns were observed in multicomponent giant unilamellar
vesicles \cite{BHW03, RKG05, VSK07, VCS08, HCC08, HVK09, SIS09, KGA11, GAF13}.
We will discuss these observations in more detail further below.

On the side of theoretical membrane science, a number of mechanisms have been
proposed that can generate micro- and nanostructures in lipid bilayers. The
purpose of the present paper is to give an overview over these mechanisms and
to explain and discuss some prominent examples. The review mainly focusses on
lipid driven mechanisms that do not involve direct protein-protein
interactions.  Since virtually all of these mechanisms are based on the phase
behavior of membranes in one way or another, the discussion will be guided by
the theory of phase transitions and critical phenomena. To make it accessible
for a general audience while still giving mathematical background, the
mathematical aspects of the discussion are presented separately in the
appendix. 

This overview is far from complete. For further information on different
aspects of the topic, the reader is referred to other recent reviews, e.g. by
Fan \etal \cite{FSH10} (focussing on nonequilibrium mechanisms), Palmieri \etal
\cite{PYB14} (focussing on mechanisms based on line active molecules), Lipowsky
\cite{Lipowsky14} (focussing on membrane shapes and membrane remodeling), and
Komura and Andelman \cite{KA14} (focussing on phase separation, phase
separation dynamics and on microemulsions). In particular, the present article
only touches on the issue of multiscale computer simulations of 
heterogeneous lipid bilayers, which is a challenge in itself and has been
discussed in several recent review articles \cite{PS09, Review09, BT13, DKP14,
BT15}. 
 
The remaining article is organized as follows: In the next section, Sec.\
\ref{sec:phase_separation}, we briefly discuss the phase behavior of lipid
membranes, focussing on lipid-lipid phase separation.  Sec.\
\ref{sec:incomplete_phs} considers mechanisms of domain formation in pure
membranes which are based on incomplete phase separation. This includes the
formation of critical clusters just above a critical demixing point (Sec.\
\ref{sec:critical}) as well as the appearance of multidomain structures on
curved vesicles that emerge in order to minimize the total bending energy
(Sec.\ \ref{sec:curved}). In Sec.\ \ref{sec:microemulsion}, we review physical
mechanisms that generate equilibrium two-dimensional microemulsions, e.g., due
to line active agents (Sec.\ \ref{sec:linactants}), or due to lipid curvature
induced elastic interactions (Sec.\ \ref{sec:bilayer_curvature} and
\ref{sec:monolayer_curvature}).  Sec.\ \ref{sec:environment} reviews
domain-stabilizing mechanisms that rely on an interaction between the membrane
and its environment, such as  membrane recycling (Sec.\ \ref{sec:recycling}),
the presence of pinning sites (Sec.\ \ref{sec:cytoskeleton}), or other
interactions that influence the phase separation kinetics (Sec.\
\ref{sec:ne_other}).  We summarize and conclude in Sec.\ \ref{sec:conclusions}.
Finally, \ref{app:gl} provides a unified mathematical description of the
domain-forming mechanisms discussed in the main text within the framework of
the Ginzburg-Landau theory.

The following abbreviations will be used in the text:
\\ -- AFM (atomic force microscopy)
\\ -- NMR (nuclear magnetic resonance)
\\ -- FRET (F\"orster resonance electron transfer)
\\ -- ESR (electron spin resonance)
\\ -- GUV (giant unilamellar vesicle)
\\ -- GMPV (giant plasma membrane vesicle)
\\ -- DPPC (dipalmitoylphosphatidylcholine)
\\ -- DMPC (dimiristoylphosphatidylcholine)
\\ -- DLPC (1,2-Dilauroyl-sn-glycero-3-phosphocholine)
\\ -- DSPC (1,2-Distearoyl-sn-glycero-3-phosphocholine)
\\ -- DOPC (1,2-Dioleoyl-sn-glycero-3-phosphocholine)
\\ -- POPC \mbox{(1-palmitoyl-2-oleoyl-sn-glycero-3-phosphocholine)}
\\ -- SM (sphingomyelin), Chol (Cholesterol) 
\\ -- {\em ld} (liquid disordered), {\em lo} (liquid ordered)

\section{Lipid phase behavior and lipid-lipid phase separation}
\label{sec:phase_separation}

The basis of lipid-driven micro- or nanodomain formation is the observation
that lipid bilayers can undergo phase transitions and in particular, that
lipids in multicomponent bilayers can phase separate. 

Already one component lipid bilayers exhibit a complex phase behavior,
which is similar for different classes of lipids \cite{Gennis89, KC94, KC98,
KC02}. At high temperatures, they are in the ''fluid'' state ($L_\alpha$),
which is characterized by a low in-plane shear viscosity and a high number of
chain defects in the hydrocarbon chains. At lower temperature, the bilayers
assume a ''gel'' state,  where the shear viscosity is higher and the
hydrocarbon chains have very few chain defects. Depending on the details of the
lipid structure, in particular the effective size and the interactions of the
polar head groups, the gel state exists in different variants: Straight chains
($L_{\beta}$ phase), collectively tilted chains ($L_{\beta'}$ phase), or even
interdigitated chains ($L_{\beta}^{\mbox{\tiny int}}$ phase).  The fluid and
the gel regions are separated by the so-called ''main'' or ''melting''
transition, which typically occurs at temperatures around room temperature or
higher. If the low temperature state is a tilted gel state, a $P_{\beta'}$
ripple phase interferes in the transition region, which corresponds to the
modulated nanostructured structure mentioned in the Introduction.  The general
structure of lipid phase diagrams seems to be quite generic and is also
reproduced by coarse-grained simulations of very simplified model ''lipid''
molecules \cite{KS05,LS07}. The most relevant phase in the biomembrane context
is the $L_\alpha$ phase.  With few exceptions \cite{PNB07}, living
organism tend to tune the lipid composition of their membranes such that 
they remain fluid.

In multicomponent lipid membranes, the lipids may not only order, but also
demix. The first experimental phase diagram of a multicomponent lipid bilayer,
a DPPC/Chol mixture, was published in 1990 by Vist and Davis
\cite{VD90}. It was derived based on deuterium NMR spectroscopy and
differential scanning calorimetry, and it included a demixed two-phase region
between two coexisting fluid phases, which were subsequently termed {\em ld}
(liquid disordered), and {\em lo} (liquid ordered) phase \cite{ST91} based on a
theoretical proposal by Ipsen et al \cite{IKM87}.  This terminology is now
widely used in descriptions of rafts.  Veatch and Keller later investigated the
same binary system by fluorescence microscopy on mixed GUVs and saw no
indication of fluid-fluid coexistence \cite{VK03,VK05}.  Around the same time,
Filippov \etal\ studied binary mixtures of DMPC and cholesterol by NMR
diffusion experiments and rationalized their findings by postulating the
existence of small ''coexisting'' {\em lo} domains in the mixtures, which
rapidly exchange lipids with an {\em ld} environment \cite{FOL03,LO09}. This
suggests an explanation for the seemingly contradictory reports on the phase
behavior of binary lipid-cholesterol mixtures \cite{VK03}: Phase separation
between {\em lo} and {\em ld} phase is typically observed with methods that are
sensitive to local ordering and rearrangements, whereas the same systems appear
homogeneous in microscopic studies that detect phase separation  on the scale
of micrometers. Hence these systems are presumably filled with very small
domains.  Indeed, recent neutron diffraction experiments by Rheinst\"adter and
coworkers provided evidence for the existence of highly ordered nanoscopic
lipid domains in the DPPC/Chol mixtures \cite{AMD13, RM13, AHS14,
TMA14}. Similar structures were observed by us in coarse-grained computer
simulations of binary mixtures \cite{TMA14, MVS13}.  Inverted nanoscale domains
({\em ld} in {\em lo}) were reported by Kim \etal\ from atomic force microscope
(AFM) imaging of DPPC/Chol monolayers\cite{KCZ13}.  

These nanodomains are clearly interesting in the raft context, but they are not
phase separated in a thermodynamic sense.  The fact that they have been
observed does not explain why such tiny structures form, and this is our key
question here which we will discuss in the following sections. 

In binary lipid membranes, fluid-fluid phase separation in a thermodynamic
sense does not seem to occur. However, it can be observed in mixtures of more
than two components.  In 2001, Dietrich \etal \cite{DBV01} reported the
formation of large phase separated fluid domains (several tens of microns) in
GUVs containing ternary mixtures of a phospholipid, cholesterol, and
sphingomyelin. Phase diagrams for a many other  ternary systems were later
studied in detail by several groups using a variety of experimental methods
\cite{VK03, VK05b, VK05, HBS05, ZWH07, HKR15, KHP15}.  Veatch and Keller
\cite{VK03} observed that fluid-fluid {\em ld}-{\em lo} separation is a
frequent phenomenon in mixtures of cholesterol, a lipid with high melting
temperature $T_m$, and a lipid with low $T_m$. Feigenson noted that both
macroscopic {\em lo}-{\em ld} phase separation and nanodomain formation can be
found in such systems, depending on the choice of lipids, and introduced the
categories of ''Type I'' and ''Type II'' mixtures to distinguish between two
types of phase behavior: The phase diagrams of type II mixtures feature a
region of macroscopic {\em ld}-{\em lo} phase separation.  This demixed region
is missing in type I mixtures and replaced by a region with prominent
nanoscopic domain formation \cite{Feigenson07, Feigenson09, KWM13}. To study
the transition between type I and type II behavior, Feigenson and coworkers
carried out systematic studies of four-component mixtures with compositions
that ''interpolate'' between typical type I and type II mixtures, and found
that ordered modulated structures emerge at intermediate compositions
\cite{KGA11, GAF13}.

Nowadays, ternary mixtures of saturated and unsaturated phospholipids (which
have a high and low melting temperature, respectively) and cholesterol serve as
standard model systems for systematic studies of raft formation in membranes
\cite{WW15, BLS16}. Large scale atomistic and coarse-grained computer
simulations have been carried out to study fluid-fluid phase separation in such
systems \cite{HH13, HH14, AF15}.  However, lipid-lipid phase separation is not
restricted to model systems, it has also been observed in membranes of natural
lipids that were extracted from brush border membranes \cite{DBV01}, and even
in membranes that were directly extracted from living cells \cite{BHS07,VCS08}.
Hence it seems to be a generic phenomenon in multicomponent membranes, and it
seems reasonable to associate it with raft formation. 

Nevertheless, the question remains why rafts are so small, and
-- possibly related -- why small nanoscopic domains are observed in model
membranes. In a classical phase separation scenario, the interface between
domains is penalized by a line tension, and the equilibrium state is one that
minimizes the length of phase boundaries, i.e., the system partitions into only
two regions {\em ld} and {\em lo}. In the following sections, we will present a
variety of mechanisms that interfere with large scale phase separation and may
lead to the formation of small domains.

\section{Domain formation due to incomplete phase separation}
\label{sec:incomplete_phs}

We first discuss isolated multicomponent membranes that exhibit macroscopic
phase separation (i.e., ''type II'' mixtures in the categorization of Feigenson
\cite{Feigenson09}). Two classes of mechanisms have been proposed that may
generate small domains in such systems: Critical fluctuations and lipid sorting
due to background curvature.

\subsection{Domains close to a critical point}
\label{sec:critical}

%  \cite{VK05} Veatch/Keller Seeing spots \\
%  \cite{VSK07} Veatch/Keller critical fluctuations in model lipid mixtures\\
%  \cite{VCS08} Veatch Critical fluctuations in plasma membrane
%  \cite{HCC08} Keller line tensions \& critical exponents in lipid membranes\\
%  \cite{HVK09} Keller Review critical phenomena in membranes\\
%  \cite{HMK12} Keller dynamic critical phenomena\\
%  \cite{SHG13} Keller coarsening dynamics
%  \cite{CH13} Connell et al critical domains with AFM
%  \cite{MVS12} Machta Critical Casimir forces theoretical and simulation\\

Keller, Veatch and coworkers proposed a strikingly simple possible explanation
why small domains are observed in membranes \cite{VK05,VSK07}: They argued that
these domains could simply be fingerprints of critical behavior in the vicinity
of critical demixing points: Even though membranes made of natural lipid
mixtures have been shown to phase separate at low temperatures, one would
expect that body temperatures are typically well above the miscibility gap.
Indeed, Veatch \etal\ studied giant plasma membrane vesicles (GMPVs) that were
extracted directly from the living rat basophil leukemia cells \cite{VCS08},
and found that they undergo a demixing transition at temperatures around $T_c
\sim 20 ^0$ C, well below the body temperature of rats (which is
similar to that of humans).  The transition temperatures for different GMPVs
where widely spread over a range of 5-10 ${}^0$ C.  Interestingly, the
lipid compositions were nevertheless always found to be near-critical. This
suggests that the small domains in real membranes may simply be critical
clusters, as occur naturally in the vicinity of critical point (see
\ref{app:gl_critical} for a mathematical description). 

\begin{figure}
\centerline{
  \includegraphics[width=0.3\textwidth]{\dir/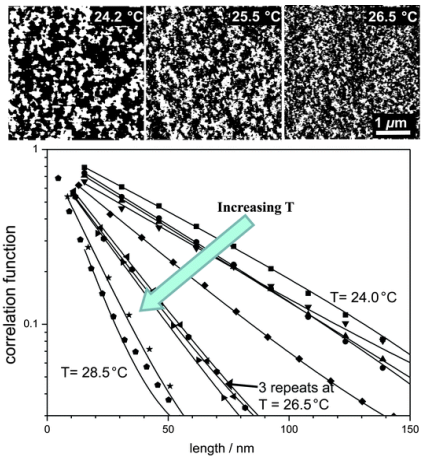}
}
\caption{
Top: AFM images of critical fluctuations in supported ternary bilayers
of SM:DOPC:Chol just above critical point (which is at $T_c=23.25{}^0$C ). 
Bottom: Semi-logarithmic plot of the corresponding radially averaged 
correlation functions, demonstrating the decrease of the correlation
length as the temperature moves away from the critical temperature.
Reprinted from S.\ D.\ Connell \etal, 
{\em Critical point fluctuations in supported lipid membranes},
Faraday Disc. {\bf 161}, 91 (2013)
\protect\cite{CH13}. Reproduced by permission of the Royal Society of 
Chemistry.
}
\label{fig:fig1}       % Give a unique label
\end{figure}

Critical clusters are characterized by a fractal structure and a broad size
distribution. The average domain size is set by the correlation length, which
diverges at the critical point. At critical compositions, the correlation
length at temperatures of 10 degrees above the demixing temperature $T_c$ can
still be of the order of 10 nm \cite{CH13}. Fig.\ \ref{fig:fig1} shows AFM
pictures of critical clusters in supported multicomponent bilayers published
along with the corresponding radially averaged correlation functions (from
Connell \etal \cite{CH13}). 

From the theory of critical phenomena, one expects several quantities to show a
peculiar power law behavior close to a critical point (see
\ref{app:gl_critical}): As one approaches the critical point at fixed critical
composition, the correlation length should diverge as $\xi \propto
|T-T_c|^{-1}$, the line tension should vanish according to $\lambda \propto
|T-T_c|$, and the width of the miscibility gap at $T<T_c$  along the tie lines
should vanish according to a power law with $\Delta c \sim |T-T_c|^{1/8}$.
These theoretical predictions have been verified by Keller and coworkers
\cite{HCC08,HVK09} using fluorescence microscopy on model multicomponent
membranes and the expected power law behavior was confirmed within the
experimental error. The same group also studied the dynamics of thermal
fluctuations close to $T_c$ \cite{HMK12} and the dynamics of phase separation
and coarsening below $T_c$ \cite{SHG13} and again found good agreement with
theoretical predictions. Connell \etal\ showed by AFM studies of supported
bilayers that the critical behavior persists up to temperatures where the
correlation length is only of the order of a few nanometers \cite{CH13}.

Hence, critical phenomena are one plausible explanation for the observation of
nanoscale raft-like structures in lipid membranes. However, they can only be
observed in membranes with close-to-critical lipid compositions. This raises
the question why living organisms should go through the effort to maintain such
highly specific compositions in their membranes. One possible answer is that it
may be of advantage to keep cell components close to phase transitions
\cite{Heimburg07}. In the vicinity of phase transitions, systems respond very
sensitively to external stimuli, and this provides efficient control mechanisms
which may be useful in the complex interaction network of a cell.  Machta \etal
\cite{MVS12} proposed that nature may also take advantage of the
relatively long-range fluctuation-driven forces between membrane inclusions
(so-called Casimir forces) that emerge close to critical points. Such forces
can possibly be exploited to manipulate the lateral organization of membrane
proteins.

\subsection{Multiple domain formation in curved geometries}
\label{sec:curved}

%  \cite{UKP09} Related, repulsive interaction between curved dimples
%    with spontaneous curvature
%\cite{Lipowsky92, Lipowsky93} First papers on domain induced budding
%\cite{JL93} domain induced budding in vesicles
%  \cite{WKA15} Andelman Komura budding of domains
%\cite{KA14} Andelman Review discusses domain induced budding
%\cite{Lipowsky14} Lipowsky review
%\cite{BHW03,BDW05} Baumgart Hess

Below the demixing point, phase separation is often coupled to conformational
changes in the membrane \cite{KA14, Lipowsky14}. One prominent example is
domain-induced budding \cite{Lipowsky92, Lipowsky93, JL93, JL96, BHW03, BDW05,
SIS09, UKP09, RUP11, WKA15}, where phase separated membrane domains bulge out
of a membrane in order to reduce the contour length of the domain boundary.
This effect was first proposed theoretically in 1992 by Lipowsky
\cite{Lipowsky92} and later confirmed experimentally by Baumgart and coworkers
\cite{BHW03, BDW05} by experiments on multicomponent vesicles.  Depending on
the membrane tension, the size of the domains, their bending stiffness, and
other elastic properties, one observes transitions between buds, dimples, and
flat domains \cite{UKP09, RUP11, WKA15}. Domain-induced budding effect does not
stabilize multidomain formation {\em per se}, but it does facilitate it by
reducing the total costs of domain nucleation.  Furthermore, the dimpled
domains repel each other through curvature-mediated interactions \cite{UKP09,
SIS09}, which slows down the aggregation.

%\cite{SCM09} experimental pipette experiments lipid sorting
%\cite{HTE09} experimental pipette experiments lipid sorting
%\cite{GGL09} Theoretical Lipowsky group multiple domain on vesicle
%\cite{HWL11} Simulation Lipowsky group multiple domains on vesicles
%\cite{AGF13} Amozon et al complex shapes on vesicles
%\cite{RMM11} CG simulations domain sorting on nm size vesicle (20 nm)

A related effect which does have the potential to induce multi-domain formation
in curved geometries is curvature-driven lipid sorting. It is based on the fact
that the coexisting phases in multicomponent bilayers often have very different
bending rigidities, e.g., membranes in the {\em lo} phase are stiffer than
membranes in the {\em ld} phase.  Pipette experiments have shown that different
lipids partition to areas of different curvature \cite{SCM09}, and that {\em
ld} phases may nucleate in regions of high local curvature \cite{HTE09}.  In
closed vesicles, it can hence be energetically favorable if the vesicle
partitions into flatter and more curved regions in order to better accommodate
the {\em lo} domains. Lipowsky and coworkers have analyzed this situation
theoretically and by computer simulations of an elastic membrane model. They
showed that the interplay of line tension and elastic energy may stabilize
conformations with multiple domains \cite{GGL09, HWL11}, provided the
coexisting phases have different bending rigidity, different Gaussian curvature
moduli, or different spontaneous curvature.  Risselada \etal \cite{RMM11}
verified this prediction with coarse-grained molecular simulations and observed
that the domain distribution on nanovesicles may change under mechanical
compression.  Feigenson and coworkers \cite{AGF13, GAF13, AF14} used computer
simulations of an elastic model to specifically study the multidomain formation
on vesicles that are constrained to almost spherical shapes. In some sense,
this corresponds to a situation where a frail rigid sheet (the {\em lo} domain)
is tightly wrapped around a rigid sphere: The {\em lo} domain breaks up into
many domains that can be much smaller than the vesicle diameter.  Gueguen \etal
\cite{GDM15} analyzed this problem analytically and established an analogy to
the theory of microemulsions.

The domain patterns observed in the simulations of Feigenson and coworkers
\cite{AGF13, GAF13, AF14} were very similar to the experimentally observed
patterns on multicomponent GUVs \cite{KGA11, GAF13} (Examples are shown in
Fig.\ \ref{fig:fig2}). However, to obtain this agreement, the authors had to
tune the bending rigidities to values that are about one order of magnitude
higher than typical experimental values and the line tensions had to be chosen
about two orders of magnitude lower than typical values reported for phase
separating lipid mixtures \cite{AGF13}. A subsequent careful analysis showed
that the simulations are strongly affected by grid renormalization effects
\cite{AF14}.  Nevertheless, the results suggest that in the experimental
systems, the curvature-driven lipid sorting mechanism must be supplemented by
another mechanism that reduces the line tension and promotes the formation of
small domains. Such mechanisms are discussed in the next section.

\section{Two-dimensional microemulsions}
\label{sec:microemulsion}

\begin{figure}
\centerline{
  \includegraphics[width=0.3\textwidth]{\dir/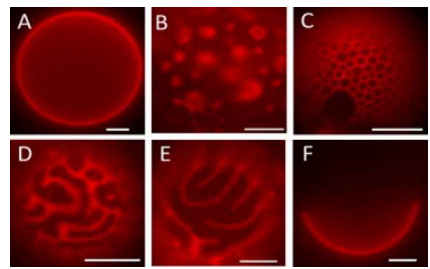}
}
\caption{
Modulated patterns on GUVs of four-component lipid mixtures composed
DSPC/(DOPC/POPC)/Chol with varying (DOPC:POPC) fractions. 
(A) Almost pure POPC (0.03:0.27); (B-E) intermediate compositions
(0.05:0.25), (0.06:0.33), (0.1;0.29), (0.05:0.25); 
(F) pure DOPC (0.3:0).  Scale bar corresponds to 10 $\mu$m.  
Reprinted from  Biophys. J. {\bf 101}, T. M. Konyakhina \etal, 
{\em Modulated phases in four-component DSPC/DOPC/POPC/Chol giant 
unilamellar vesicles}, L06--L10 \protect\cite{KGA11}.
Copyright (2013) with permission from Elsevier.
}
\label{fig:fig2}       % Give a unique label
\end{figure}

The domain formation mechanisms discussed in the previous section rely on the
existence of true fluid-fluid phase separation.  They cannot explain the
observations of equilibrium  nanoclusters in model membranes that do not
exhibit a miscibility gap (see Sec.\ \ref{sec:phase_separation}).  There also
exist a number of mechanisms that stabilize small domains at the level of the
equilibrium phase diagram. Lipid mixtures that show such a behavior typically
have a well-defined characteristic wave length, and are called
''microemulsions'' in the language of the theory of phase transitions
\cite{GompperSchick94}.  This section is devoted to describing mechanisms that
can produce such microemulsions in two dimensions. A mathematical framework for
the description of these structures is provided in \ref{app:gl_microemulsion}.

\begin{figure}
\centerline{
  \includegraphics[width=0.4\textwidth]{\dir/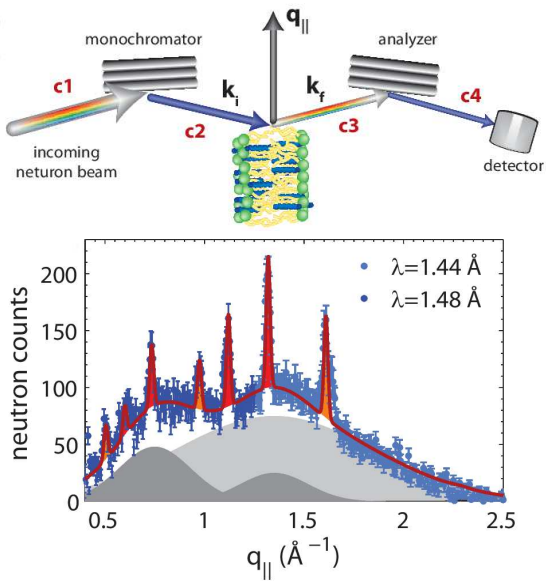}
}
\caption{
Experimental evidence for the existence of nanoscale {\em lo} clusters in
binary DPPC-cholesterol mixtures: Top: In-plane neutron diffraction pattern
measured on a solid-supported stack of highly oriented DPPC membranes
containing 32.5 \% deuterated cholesterol.  The peaks correspond to small
highly ordered clusters of {\em lo} phase, and the broad background to the
surrounding sea of {\em ld} phase.  Bottom: Corresponding experimental setup.
Reproduced from Toppozino \etal \protect\cite{TMA14}.
}
\label{fig:fig3}       % Give a unique label
\end{figure}

The experiments on model membranes suggest that several types of domains with
different characteristic length scales may exist and possibly coexist in such
membranes.  On the one hand, large modulated micron-size patterns have been
observed in GUVs by fluorescence microscopy \cite{BHW03, RKG05, SIS09, KGA11,
GAF13}.  An example is shown in Fig.\ \ref{fig:fig2}.  On the other hand, as
already reviewed in Sec.  \ref{sec:phase_separation}, increasing experimental
evidence points to the existence of nanosize clusters in certain multicomponent
mixtures, e.g., in binary phospholipid-cholesterol mixtures or in those
multicomponent mixtures that Feigenson \cite{Feigenson09}  classified as ''type
I'' mixtures.  Fig.\ \ref{fig:fig3} shows a recent example of an in-plane
neutron diffraction pattern for a mixture of DPPC and deuterated cholesterol,
which clearly shows a series of weak Bragg peaks superimposed to a broad
fluid-like spectrum \cite{TMA14}.  This is attributed to the existence of small
ordered cholesterol-rich {\em lo} clusters in a sea of disordered {\em ld}
phase.  Recent all-atom simulations by Sodt \etal \cite{SSG14} indicate that 
even in phase-separating mixtures, the {\em lo} phase itself may have a 
nanoscale substructure.

One factor which has the potential of stabilizing finite size domains in phase
separating systems is electrostatics \cite{LQG05,Travesset06}. Modulated
structures may emerge due to the competition of line tension and and
electrostatic dipolar interactions between the head groups. This mechanism is
known to be effective in monolayers at the air-water interface, where it leads
to the formation of complex micron-size domain patterns. In contrast to
monolayers, however, bilayers are fully immersed in water, where the strength
of electrostatic interactions is significantly weakened due to the high
dielectric permittivity. Moreover, the interactions are largely screened at
physiological salt concentrations. Hence electrostatic effects should be
much weaker than in monolayer systems. Amazon and Feigenson recently studied
the effect of electrostatic dipolar interactions on domain formation in
vesicles by coarse-grained simulations and concluded that electrostatic dipolar
interactions might suppress phase separation and stabilize domains of sizes in
the order of 10 nm in model membranes with pure water environment (no salt),
but not under physiological conditions \cite{AF14}.

However, this study treated the membranes as infinitely thin surfaces
in space. In reality, they have finite thickness, and their hydrophobic
core has a low dielectric constant and does not contain ions.  Liu \etal
\cite{LQG05} argued that long-ranged electrostatic interactions could be
transmitted across the interior of the membrane.  They developed an analytical
theory for this problem and predicted that electrostatics  may stabilize
domains of submicron sizes (up to around $\sim$ 100 nm). Travesset
\cite{Travesset06} independently considered the same problem and came to the
rather different conclusion that electrostatic interactions cannot stabilize
domains that are much larger than the Debye length in the surrounding medium,
which is around 1 nm at physiological conditions. Both studies used similar
approximations (most notably, the linear Poisson-Boltzmann approximations), but
they differ in their Ansatz where to place the dipoles. Liu \etal\ considered
dipoles that are buried inside the membrane, whereas Travesset assumed that the
dipoles are located at the membrane surface. Hence we can conclude that
electrostatic interactions between dipolar membrane components can disrupt
phase separation and generate microemulsion structures in multicomponent
membranes, but only if the dipoles are forced to reside inside the membrane.
Since the apolar membrane interior tends to repel polar and charged monomers,
this does probably not happen very frequently, and electrostatic interactions
can presumably be neglected in most cases.

\begin{figure}
\centerline{
  \includegraphics[width=0.3\textwidth]{\dir/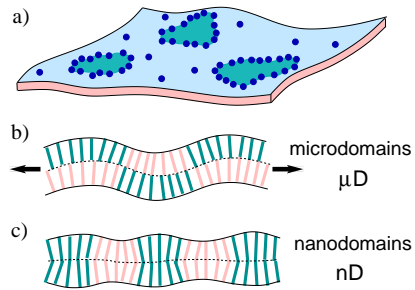}
}
\caption{
Mechanisms that can stabilize two dimensional microemulsions.
(a) line active molecules, discussed in Sec.\ \ref{sec:linactants},
leading to domains on length scales in the range of a few nanometers;
(b) bilayer curvature coupling (Leibler-Andelman mechanism), discussed 
in Sec.\ \ref{sec:bilayer_curvature}
and leading to domains on length scales of the order 100 nm to micrometers;
(c) monolayer curvature coupling, discussed in Sec.\
\ref{sec:monolayer_curvature} and leading to domains on length scales 
of the order 10 nm. 
}
\label{fig:fig4}       % Give a unique label
\end{figure}

In the following, we will mainly focus on the mechanisms sketched in Fig.\
\ref{fig:fig4}: Mechanisms that stabilize microemulsions due to line active
membrane components, or due to a coupling between the local lipid composition
and the local bilayer or monolayer curvature.

% \cite{SSG14} Sodt ternary DOPC/DPPC/chol - substructure in lo phase\\

\subsection{Linactants}
\label{sec:linactants}

\begin{figure}
\centerline{
  \includegraphics[width=0.3\textwidth]{\dir/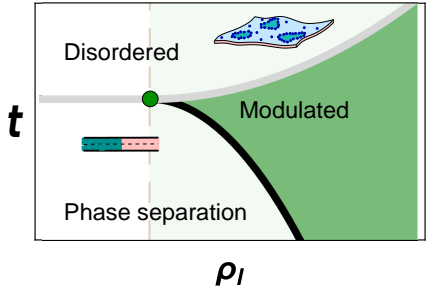}
}
\caption{
Generic mean field phase diagram for ternary mixtures with linactants in the
plane of the temperature-like  parameter $t$ vs.\ bulk linactant concentration
$\rho_l$ (i.e.\ , linactant concentration far from domain boundaries), as
calculated in \ref{app:gl_linactant}.  The relation between $t$, the
temperature $T$, and the surfactant concentration $\rho_l$ is roughly linear,
$T = T_c + \alpha (t - \nu \rho_l)$ where $T_c$ is the demixing temperature of
the pure system and $\alpha, \nu$ are phenomenological constants.
The light lines denote second order phase transition, the 
dark lines first order phase transition, the circle indicates the position 
of a Lifshitz multicritical point (see text). The thin dashed line corresponds
to the Lifshitz line, which marks the onset of disordered structure formation 
in the homogeneous phases \protect\cite{GompperSchick94}.  }
\label{fig:fig5}       % Give a unique label \end{figure}
\end{figure}

Adding surfactants is an efficient way of stabilizing microemulsions
\cite{GompperSchick94}. Safran, Andelman and coworkers \cite{HKA09, HKA12,
BPS09,YBS10,YS11,PS13,PGS14,PYB14} have argued that the heterogeneities
observed in biomembranes might to a large extent be attributed to the presence
of line active molecules.  An extensive recent review with many references can
be found in \cite{PYB14}, here we describe only some basic aspects of the
mechanism. Line active molecules (linactants) aggregate to domain boundaries,
thereby reducing the line tension, until they may eventually destroy the
demixing transition and stabilize ordered or disordered modulated structures.
A simple generic mechanism is outlined in \ref{app:gl_linactant} and yields the
phase diagram shown in Fig.\ \ref{fig:fig5}. This calculation is very much
simplified and does not account for important effects such as, e.g., the fact
that domain boundaries eventually become saturated with linactants.
Nevertheless, it qualitatively reproduces the basic structure of the expected
mean-field phase behavior of systems containing linactant molecules. Here the
term ''mean-field'' refers to the approximation that long range thermal
fluctuations, e.g., fluctuations of the domain boundaries, are neglected.

At low linactant concentrations, the main effect of linactants is to reduce the
line tension. The system still undergoes a demixing transition, but the
transition point is shifted to lower temperatures.  If the linactant
concentration reaches a certain threshold value, the line tension vanishes,
demixing transition is suppressed and an ordered modulated phase forms instead.
Beyond this threshold, the homogeneous phase also acquires local structure with
a well-defined characteristic wave length, which is of the order of the
underlying molecular correlation length, i.e., in the range of a few
nanometers.  This is the structure called ''microemulsion''. The crossover from
a fully homogeneous state to the microemulsion state is marked by the so-called
''Lifshitz line''. The demixing line meets the order-disorder line, i.e.,
the transition line between the microemulsion and the modulated state,
at a so-called ''Lifshitz multicritical point'' (see
\ref{app:gl_microemulsion}). The characteristic wave length diverges at
the Lifshitz point.

As discussed in \ref{app:gl_linactant}, the mean-field scenario is modified in
the presence of thermal fluctuations. The Lifshitz point disappears, the
order-disorder transition shifts to lower temperatures and becomes first order,
and in return, the disordered ''microemulsion'' widens significantly. The
characteristic wave length of the microstructures no longer diverges at the
transition to the two phase region, but stays finite.

Safran and coworkers \cite{BPS09,YBS10,YS11,PS13,PGS14,PYB14} 
and Andelman and coworkers \cite{HKA09, HKA12} argued that this
linactant scenario may partly account for the heterogeneities that have been
observed in membrane systems. In particular, they suggested that hybrid lipids
with one saturated and one unsaturated tail may serve as linactants that
accumulate at {\em ld}-{\em lo} boundaries and reduce the line tensions. This
hypothesis is consistent with a number of experimental findings \cite{SSS11,
KGA11, GAF13}. Here we specifically discuss the work on four-component mixtures
by Feigenson and coworkers \cite{KGA11, GAF13} which lead to the
structures shown in Fig.\ \ref{fig:fig2}. 

These structures were obtained within a systematic study of four-component
mixtures  with compositions chosen such that they interpolate between the phase
separating ''type II'' mixture DSPC/DOPC/Chol (Fig.\ \ref{fig:fig2}F) and the
''type I'' mixture DSPC/POPC/Chol, which does not phase separate \cite{KGA11}.
If one starts with DSPC/DOPC/Chol and gradually substitutes the fully
unsaturated lipid DOPC with the hybrid lipid POPC, phase separation gives way
to the modulated structures shown in Fig.\ \ref{fig:fig2} B-E. This suggests
that hybrid lipids act as linactants and reduce the line tension, thus
facilitating one of the domain forming mechanisms that can generate large
micron size structures (such as the lipid sorting mechanism described in Sec.\
\ref{sec:curved} or the Leibler-Andelman mechanism to be described in Sec.\
\ref{sec:bilayer_curvature}). Indeed, molecular simulations of a similar
mixture have indicated that the substitution of fully saturated lipids by
hybrid lipids leads to a decrease of the line tension between coexisting phases
\cite{AF15}. 

Intriguingly, experimental studies of a four-component mixture consisting of
purely nonhybrid lipids revealed a scenario which is very similar to that shown
in Fig.\ \ref{fig:fig2} \cite{HDG13}.  In that work, DOPC was not substituted
with the hybrid lipid POPC, but with the fully saturated lipid DLPC.
%DLPC has a shorter chain length than DSPC and therefore its melting point is
%lower \cite{KC98}, which turned out to be sufficient to produce a type I
%mixture, but not sufficient to produce a phase separating type II mixture. 
Thus lipids do not need to be hybrid lipids to reduce the line tension.
Nevertheless, they may still act as linactants, if they accumulate at domain
boundaries. According to the general description of linactants outlined in
\ref{app:gl_linactant}), all molecules and particles \cite{Cheung14} that are
attracted to domain boundaries have linactant properties.

According to the scenario displayed in Fig.\ \ref{fig:fig5}, linactants at high
concentrations might altogether suppress phase separation, which suggests that
the linactant mechanism might be responsible for the lack of global phase
separation in the type I mixtures DSPC/POPC/Chol or
DSPC/DLPC/Chol.  However, this would imply that the pure
DSPC/Chol mixture, without linactant, should  phase separate into 
a {\em lo} and {\em ld} phase, which is not the case (see Sec.\
\ref{sec:phase_separation}). Furthermore, neutron scattering studies on small
four-component vesicles made of DSPC/(DOPC/POPC)/Chol mixtures with low DOPC
content (still type I mixtures) have indicated that the nanodomain size regime
is mainly controlled by the bilayer thickness mismatch between the nanodomains
and the surrounding phase \cite{HPP13}, which suggests that linactant effects
are not dominant in the type I regime. Hence the situation seems to be more
complicated than suggested by the simple linactant scenario.  We will now
discuss other mechanisms that can prevent phase separation and stabilize
nanoscale clusters.

%  \cite{BPS09,YBS10,YS11} Brewster, Yamamoto, Safran linactive \\
%  \cite{PS13,PGS14} Palmieri hybrid lipid\\
%  \cite{PYB14} * Palmieri/Safran review 2014 (line active)\\
%  \cite{Cheung14} simulation example (linactive nanoparticles)\\
%  \cite{SSS11} experimental example where hybrid lipids induce mixing
%  \cite{KGA11, GAF13} Feigenson ternary mixture with hybrid lipids
%   large scale modulations at intermediate concentrations
%   theory \cite{AGF13} - bending modulus and Gaussian modulus 
%    couples to curvature. Factor 10 difference between lo (strong) and ld
%    - ME-like structures on vesicles
%  \cite{AF15} Feigenson phase separation and domain formation
%    in 4comp mixture (simulation) hybrid lipid mechanism
%  \cite{AF15} Feigenson phase separation and domain formation
%    in 4comp mixture (simulation) hybrid lipid mechanism

\subsection{Bilayer curvature coupling: The Leibler-Andelman mechanism}
\label{sec:bilayer_curvature}

\begin{figure}
\centerline{
  \includegraphics[width=0.3\textwidth]{\dir/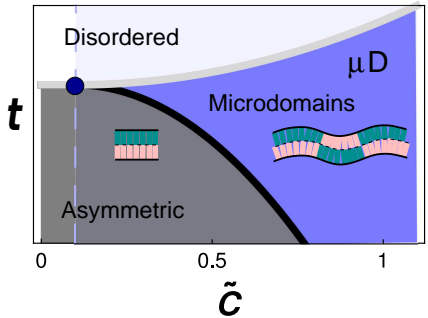}
}
\caption{
Generic mean-field phase diagram for a symmetric multicomponent bilayer 
(here ''symmetric'' means that both leaflets are equivalent and can freely 
exchange lipids) subject to the Leibler-Andelman mechanism in the plane of
temperature-like parameter $t$ vs.\ dimensionless curvature coupling parameter
$\tilde{C}$.  It has the same topology than the phase diagram of Fig.\
\protect\ref{fig:fig5}.  The light lines correspond to second order phase
transitions, the dark lines to first order phase transitions, the thin blue
dashed line to a Lifshitz line, and the circle to the position of the Lifshitz
point, where the wavelength of modulations diverges. Redrawn from
\protect\cite{BS16} with slightly different parameters (see
\ref{app:gl_curvature}). 
}
\label{fig:fig6}       % Give a unique label
\end{figure}

Already in 1987, Leibler and Andelman \cite{LA87} proposed a mechanism that may
stabilize modulated structures and microemulsions with typical sizes in the
range of 100 nm to micrometers in mixed membranes.  The basic idea is sketched
in Fig.\ \ref{fig:fig4}b). If the local lipid compositions on two apposing
monolayer leaflets differ from each other, the bilayer acquires ''spontaneous
curvature'', i.e., a tendency to bend in a certain direction \cite{PS11}. An
unconstrained bilayer would simply curve around and close up to form a vesicle
\cite{SPA90}.  By applying lateral tension, the membrane can be forced to
remain planar on average.  Now, if the lipids in each monolayer have a tendency
to demix, the membrane can accommodate to the curvature stress by developing
lateral structure, such that regions with different spontaneous bilayer
curvature alternate with each other.  For strong demixing force, this mechanism
generates a variety of competing ordered mesophases with striped or hexagonal
domain patterns \cite{LA87, HM94, KGL99, GDM15}. However, membranes may develop
fluctuating lateral heterogeneities with a well-defined characteristic size
even in the mixed regime. This was noted by Liu \etal \cite{LQG05} and more
recently by Schick \cite{Schick12}, and Schick and coworkers analyzed the
possible implications for raft formations in some detail \cite{SS13, SMS14,
GS16}. The characteristic wave length of the mesostructures is of order 100 nm
to $\mu$m or higher for realistic membrane parameters (see
\ref{app:gl_curvature}) and scale as $1/\sqrt{\Gamma}$ with the membrane
tension $\Gamma$ \cite{LQG05, Schick12}. 

Shlomovitz \etal \cite{SS13,SMS14} pointed out that the Leibler-Andelman
mechanism induces a Lifshitz point scenario which is very similar to that
expected for linactants (Fig.\ \ref{fig:fig5}), if one replaces the linactant
concentration $\rho_l$ by the strength of the curvature-composition coupling
difference $\tilde{C}$. The underlying mathematical description is outlined in
\ref{app:gl_curvature}.  It results in the mean-field phase diagram shown in
Fig.\ \ref{fig:fig6} (redrawn from Ref.\ \cite{BS16}, see also \cite{SMS14}),
which is very similar to Fig.\ \ref{fig:fig5}. At weak curvature coupling, the
lipids phase separate and the two phases partition to the two sides of the
monolayer. For strong coupling, a modulated structure develops. The two regimes
are separated by a Lifshitz multicritical point.  As discussed earlier and in
\ref{app:gl_microemulsion}, the order/disorder transition between the
disordered fluid and the ordered microdomains becomes first order in the
presence of thermal fluctuations, and the Lifshitz point is destroyed.
Simulations by Shlomovitz \etal \cite{SMS14} and Sadeghi \etal \cite{SMV14}
indicate that it is replaced by a tricritical point: If one increases the
curvature coupling, the demixing transition first becomes first order before
being replace by the order/disorder transition.

%  \cite{Schick12} Schick \\
%  \cite{SS13} Shlomovitz Schick asymmetric\\
%  \cite{SMS14} Shlomovitz Schick phase diagram (+simulations)
%  \cite{HWL11} Simulation, domains on vesicles w. different bilayer curvature
%  \cite{GDM15} Gueguen, theory, vesicles with lipid sorting & bilayer curv.
%  \cite{BHW03} Baumgart Hess \\
%  \cite{RKG05} modulated phases on GUVs with sphingomyelin, Chol, DOPC
%  \cite{HM94} Harden McKintosh (strong segregation)
%  \cite{KGL99} Kumar et al phase diagram \\
%  \cite{LQG05} Discusses this possibility for rafts.
%    Argues that 330 $\mu$ m${}^{-1}$ are necessary to generate 100 nm rafts
%  \cite{LA87} Leibler Andelman 
%  \cite{SPA90} Safran Pincus vesicle formation in surfactant mixtures 
%  \cite{SMV14} Sadeghi Mueller Vink simulation \\
%  \cite{CK08} Collins/Keller Coupling of phases across membranes \\
%  \cite{KWT09} Kiessling domain coupling in asymmetric membranes (review)\\
%  \cite{NSC15} Review Nickels bilayer asymmetry and cross-bilayer coupling, 
%  \cite{BS16} Leonie phase diagrams.
%  \cite{SS13} Shlomovitz Schick asymmetric.
%  \cite{HMD16} Katsaras ... control of domains through interleaflet coupling
%  \cite{CK08} Collins ordering on one side induces domains on the other

The Leibler-Andelman mechanism requires the lipid composition on the two
monolayers to be different, hence it seems to rely on the assumption that the
lipid domains on opposing monolayers are anticorrelated or at least decoupled
\cite{FMS13}.  Experiments and simulations rather suggest that lipid domains
across bilayers tend to be in registry \cite{KWT09, BHH15, NSC15}.  Collins and
Keller \cite{CK08} studied asymmetric lipid bilayers, with lipid compositions
chosen such that only one side is phase separating, by fluorescence microscopy.
They found that domain formation on the phase separating side may induce domain
formation on the other side.  Conversely, they also reported that the coupling
to the non phase separating leaflet may suppress domain formation on the phase
separating leaflet. Recent neutron scattering experiments by Heberle \etal
\cite{HMD16} showed a similar coupling on the subnanometer scale: A disordered
fluid leaflet can partially fluidize an ordered apposing leaflet. Hence the
domain structures on the two sides of bilayers seem to be rather strongly
coupled.

Nevertheless, the Leibler-Andelman mechanism may still be effective in bilayers
with positive domain coupling for several reasons. First, the energy gain
associated with bilayer curvature driven microdomain formation may compensate
the energy costs of staggered domain arrangements even in membranes with
positive domain coupling \cite{BS16}.  Second, Williamson and Olmsted
\cite{WO15} have recently shown by computer simulations that bilayers may be
kinetically trapped in a configuration with anticorrelated (staggered) domains
even if the thermodynamically favored state is one where domains are in
registry. Third, real biomembranes are typically asymmetric. Even domains that
are in perfect registry usually have different lipid compositions on both sides
and hence the membrane can be expected to develop spontaneous bilayer
curvature, which will be different for different phases. This is sufficient to
put the Leibler-Andelman mechanism to action.  Shlomovitz and Schick
\cite{SS13} studied specifically the case where only one side of a bilayer has
a driving force towards phase separation, and argued that this is sufficient to
bring about domain formation on both sides through a phase coupling of domains
across the bilayer. 

The Leibler-Andelman mechanism is one of the candidates that may explain the
observation of modulated micron or submicron structures in multicomponent GUVs
\cite{BHW03, RKG05, SIS09, KGA11, GAF13}. In Sec.\ \ref{sec:curved}, we have
already discussed an alternative mechanism, the lipid sorting mechanism, which
may account for multidomain formation in close-to-spherical vesicles. On the
other hand, Baumgart \etal \cite{BHW03} and Rozovsky \etal \cite{RKG05} have
observed micron size structures in vesicles that are clearly not spherical. The
relation between bilayer curvature induced structure formation and lipid
sorting induced structure formation has been discussed theoretically by Hu
\etal \cite{HWL11} and by Gueguen \etal \cite{GDM15}. Apart from the fact that
the lipid sorting mechanism requires background curvature and does not work for
planar membranes, the two mechanisms are quite similar. In particular, the
theory of lipid sorting induced structure formation can formally be mapped onto
a theory of microemulsions on curved geometries \cite{GDM15}.

%  \cite{WKA15} Andelman Komura budding of domains

\subsection{Monolayer curvature coupling}
\label{sec:monolayer_curvature}

%  \cite{Berkowitz09} Simulations of phospholipids + chol (review)
%  \cite{ZLS15} Zhang MARTINI Simulations of DPPC/chol 
%  \cite{WTE12} Waheed MARTINI Simulations DPPC/chol 
%  \cite{DAA15} MARTINI Simulations of phospholipids/chol  (DPPC,POPC,DOPC,DUPC)
%    (still good solvent)

%  \cite{LS05,SDL07} our model
%  \cite{WBS09} our model: elastic properties
%  \cite{WS10} our model: signature of curvature instability 

%  \cite{MVS13} FS: rafts in simulations
%  \cite{SDL14} FS: More on theory
%  \cite{BS16} Leonie phase diagrams
%  \cite{TMA14} Laura clusters in DPPC/chol + more simulations

%  \cite{KAC05} Kuzmin et al monolayer curvature reduces line tension
%  \cite{BM13} Mouritsen review fluid mosaic- stresses role of curvature
%
%  \cite{KHR13} Kollmitzer monolayer curvature of raft-forming lipids
%  \cite{VRP14} Mueller simulation method to determine spontaneous curvature
%    of monolayer in a binary mixture (implicit solvent CG model and MARTINI)

\begin{figure}
\centerline{
  \includegraphics[width=0.3\textwidth]{\dir/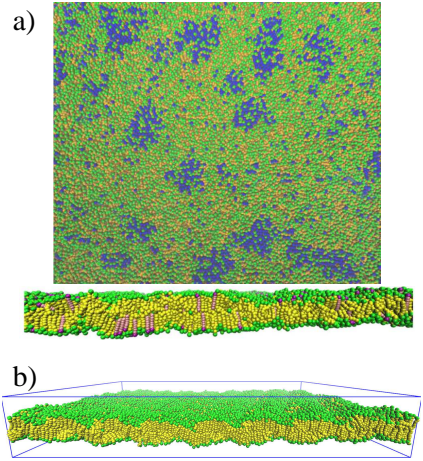}
}
\caption{
(a) Top view and side view of small clusters in simulations of tensionless
binary membranes. ''Cholesterol'' molecules are blue and pink, 
''phospholipid'' molecules are green and yellow.
(From S.\ Meinhardt, PhD Thesis \protect\cite{Meinhardt_Thesis},
similar configurations can be found in \cite{MVS13} and \cite{TMA14}.)
Panel (b) shows for comparison a ripple state configuration
($P_{\beta'}$) in a one-component membranes, which is a modulated structure
with a similar length scale (Reproduced from \protect\cite{SDL14}).
}
\label{fig:fig7}       % Give a unique label
\end{figure}

The mechanisms discussed so far still do not explain the observation of
nanoscale clusters in binary lipid mixtures, e.g., DPPC/Chol mixtures,
which were discussed in Sec.\ \ref{sec:phase_separation}.
These mixtures do not phase separate on a global scale, which rules out an
interpretation in terms of critical clusters, and they clearly do not contain
linactant components. Comparable clusters are also observed in planar
tensionless membranes in computer simulations \cite{MVS13, TMA14,
Meinhardt_Thesis}, which rules out interpretations in terms of the lipid
sorting mechanism and the Leibler-Andelman mechanism. Fig.\
\ref{fig:fig7} shows configuration snapshots of such structures which were
obtained from semi-grandcanonical Monte Carlo simulations of a coarse-grained
model for binary lipid bilayers \cite{LS05, SDL07, MVS13}.  In these
simulations, the overall number of molecules was kept constant, but the
composition was allowed to fluctuate, and all molecules could at any time
switch their identity from ''lipid'' to ''cholesterol''. This rules out 
the possibility that the clusters simply result from
incomplete phase separation. 

\begin{figure}
\centerline{
  \includegraphics[width=0.3\textwidth]{\dir/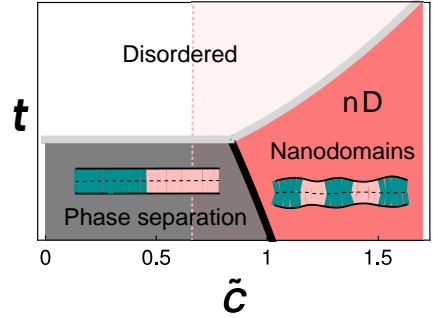}
}
\caption{
Generic mean-field phase diagram for a symmetric multicomponent bilayer
subject to the monolayer curvature coupling mechanism in the plane of 
temperature-like parameter $t$ vs.\ dimensionless curvature coupling 
parameter $\tilde{C}$. The light lines correspond to second order phase 
transitions, the dark lines to first order phase transitions, and the thin
red dotted line to a Lifshitz line.  The order/disorder line and the 
demixing line meet in a multicritical point. In contrast to 
Figs.\ \protect\ref{fig:fig5} and \protect\ref{fig:fig6}, 
the wavelength of modulations remains finite and small here.
Redrawn from \protect\cite{BS16}.
}
\label{fig:fig8}       % Give a unique label
\end{figure}

An analysis of the domain patterns showed that they have a characteristic
length scale of around 25 nm, which is not only the order of magnitude that has
been discussed for rafts, but also comparable to the periodic wave length of
the modulated ripple phase $P_{\beta'}$ \cite{KC98, LS07} discussed in Sec.\
\ref{sec:phase_separation} (a typical configuration is shown in Fig.\
\ref{fig:fig7}, bottom), and to the wavelength of a soft peristaltic mode which
we have observed in the spectrum of the $L_{\beta'}$ gel phase \cite{WS10} in
one-component membranes. These observations lead us to conjecture a connection
between the nanodomains in mixed membranes and the ripple structure in
one-component membranes, and to propose a domain forming mechanism which is
driven by a coupling of the local lipid structure to the monolayer curvature.
The fact that internal curvature stress in membranes should be taken into
account in theories for rafts had also been stressed by Bagatolli and Mouritsen
in a recent review \cite{BM13}. 

The basic picture behind the monolayer curvature coupling mechanism is sketched
in Fig.\ \ref{fig:fig4}c). We describe bilayers in terms of two coupled
elastic monolayers.  This idea goes back to Dan \etal \cite{DPS93}, and we have
recently verified the validity of the approach at the example of copolymeric
bilayers \cite{LPS13}. Furthermore, we assume that the lipid composition
couples to the spontaneous curvature of monolayers -- i.e., monolayers have a
tendency to bend inwards or outwards depending on their local composition.
Since monolayers are coupled to each other, this creates internal elastic
stress, which can be partially relieved at domain boundaries \cite{MVS13,
SDL14}. Kuzmin \etal \cite{KAC05} showed already in 2005 that a mismatch of
spontaneous monolayer curvature in coexisting lipid phases can lead to a
reduction of the line tension. If the curvature stress is large, the system
reacts by keeping domain sizes small \cite{MVS13, SDL14, BS16}.  

In practice, determining monolayer curvatures from experiments or simulations
is a difficult task. In simulations, they are typically calculated from the
first moment of the pressure profile \cite{Safran94} and the results are
plagued by large statistical errors.  Barrag\'an Vidal \etal \cite{VRP14} have
recently proposed an alternative method which allows to extract directly the
strength of curvature coupling, and applied it to mixtures of DPPC with various
other lipids, but unfortunately not yet to cholesterol. On the experimental
side, Kollmitzer \etal \cite{KHR13} recently estimated values for the
spontaneous curvature of various lipid mixtures from neutron scattering data on
inverted hexagonal lipid phases based on simple mixing rules. Their results
suggest that the concentration of cholesterol should indeed have a strong
influence on the spontaneous curvature of lipid-cholesterol mixtures, as
assumed by the monolayer curvature coupling model.

The mathematical background of the theory is briefly summarized in
\ref{app:gl_curvature}. In the symmetric case (two equivalent monolayers and
critical lipid composition), the theory yields the mean-field phase diagram
shown in Fig.\ \ref{fig:fig8}. It has obvious similarities to the phase
diagrams discussed previously, Figs.\ \ref{fig:fig5} and \ref{fig:fig6}. The
phase behavior is again controlled by the strength of curvature coupling,
$\tilde{C}$. At low coupling, the system phase separates, and at high coupling,
nanodomains form. However, in contrast to Figs.\ \ref{fig:fig5} and
\ref{fig:fig6}, the two regimes are not connected by a Lifshitz point. 
The characteristic wave length remains
finite for all values of the renormalized curvature coupling $\tilde{C}$ down
to the Lifshitz line. It depends on the elastic parameters of the membrane and
is of the order of the membrane thickness, i.e., a few nanometers. 

The same theory can also be used to describe the formation of the ripple
structure $P_{\beta'}$ (in a very simplified manner), since it does not
explicitly require that the two coexisting phases have different lipid
compositions. They may also simply have different order, i.e., gel and fluid
order. Provided that the gel and fluid state have largely different spontaneous
monolayer curvature, our theory would hence also predict the existence of
modulated structures or phases in the vicinity of the main transition -- in
agreement with the experimental observation that the ripple phase seems to be a
generic phenomenon in one-component membranes with a tilted (i.e.\, elastically
stressed) gel phase $L_{\beta'}$ \cite{KC98, KC02}.

Reigada and Mikhailov \cite{RM16} have recently carried out a linear stability
analysis and dynamic Ginzburg-Landau simulations of a monolayer curvature
coupling model and calculated a phase diagram in a different plane of
parameters than Fig.\ \ref{fig:fig8}. They found lamellar, hexagonal and phase
separated structures. Interestingly, they observed that even phase separating
systems may initially be trapped in a nanostructured state after a quench from
the disordered phase. Hence it may be possible to observe kinetically
stabilized nanoscale domains even in parameter regions where they are not
thermodynamically stable.

\begin{figure}
\centerline{
  \includegraphics[width=0.3\textwidth]{\dir/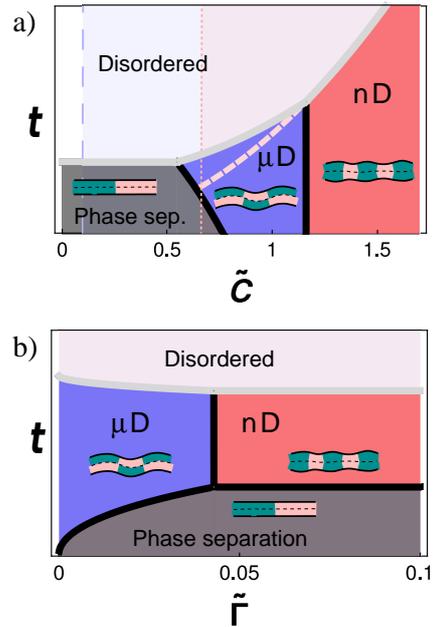}
}
\caption{
Examples of complex phase diagrams resulting from the competition of
the Leibler-Andelman mechanism and the monolayer curvature coupling mechanism.
(a) Phase diagram in the plane of the temperature-like parameter $t$
vs.\ dimensionless curvature coupling parameter $\tilde{C}$ for 
weakly positively coupled bilayers.
(b) Phase diagram in the plane of the temperature-like parameter $t$ 
vs.\ dimensionless membrane tension parameter $\tilde{\Gamma}$
at fixed curvature coupling parameter, which was chosen such that
the stable phase at low tension is the microdomain phase.
The light solid lines denote second order phase transitions, 
the dark lines first order transitions;
the thin blue dashed line the Lifshitz line with respect to microdomains,
the thin red dotted line the Lifshitz line with respect to nanodomains.
The thick pink line denotes a line below which nanodomains may form
within the $\mu D$ regime close to microdomain boundaries.
Redrawn from \protect\cite{BS16} with slightly different parameters,
see \ref{app:gl_curvature}.
}
\label{fig:fig9}       % Give a unique label
\end{figure}

Since the monolayer curvature coupling mechanism relies on a coupling between
lipid composition and curvature coupling, it is bound to compete with the
Leibler-Andelman mechanism described in Sec.\ \ref{sec:bilayer_curvature}.
Whenever monolayer curvature coupling is possible, bilayer curvature coupling
becomes possible as well. We have recently used the unified mathematical
description outlined in \ref{app:gl_curvature} to study the interplay of the
two mechanisms \cite{BS16}. Two selected phase diagrams are shown in Fig.\
\ref{fig:fig9}. Whereas the disordered micro- and nanostructures can easily
coexist (as long as the curvature coupling parameter $\tilde{C}$ is higher than
the respective Lifshitz lines), the ordered structures are found to exclude
each other, and microstructured modulated phases are separated from
nanostructured modulated phases by first order phase transitions. 

Interestingly, the theoretical phase diagram of Fig.\ \ref{fig:fig9}a) features
the same sequence of morphologies from nanodomains {\em via} microdomains to
phase separation than the experimental four-component model membrane systems
studied by Feigenson \etal\ (Fig.\ \ref{fig:fig2}). One should note that the
dimensionless coupling parameter $\tilde{C}$ in fact depends on a combination
of elastic parameters, $\tilde{C} \propto \hat{c} \sqrt{k_c/g}$, where
$\hat{c}$ is the ''bare'' coupling between lipid composition and monolayer
curvature, $k_c$ is the bending rigidity of the membrane, and $g$ is
proportional to the free energy penalty on domain boundaries, i.e., the ''local
line tension'' (see \ref{app:gl_curvature}). Variations of any of these
parameters can trigger a morphological transition. A sequence of discontinuous
transitions such as those observed in the experimental system of Fig.\
\ref{fig:fig2} could be brought about if the gradual substitution of DOPC with
POPC had at least one of the following effects: {\em Either} a gradual increase
of the (bare) coupling between lipid composition and spontaneous monolayer
curvature, {\em or} a gradual increase of the average bending rigidity of the
bilayer, {\em or} a gradual decrease of the ''local line tension'' at domain
boundaries. At least the line tension seems to show this dependence
\cite{Feigenson_private} (see also Sec.\ \ref{sec:linactants}). Hence the
theoretical scenario of Fig.\ \ref{fig:fig9}a) might account for the
discontinuous transitions between nano- and microscale patterns and global
phase separation observed in four-component membranes. 

For other coupled microemulsions, the combination of ordering mechanisms often
results in  complex structures that combine elements of both underlying
patterns (e.g.\ Hirose \etal \cite{HKA09, HKA12}).  Here, we find that
different curvature coupling induced patterns tend to suppress each other.
This is because the competing structures are associated with different
correlations between domains across the bilayers.  In the microdomain phase
stabilized by the Leibler-Andelman mechanism, they are staggered, and in the
nanodomain phase stabilized by the monolayer coupling mechanism, they are in
registry.  Therefore, the two structures cannot easily be combined.
Nevertheless, membranes in a microstructured phase may still feature
nanodomains at the boundaries of the microdomains (below the thick pink
line in Fig.\ \ref{fig:fig9}a) \cite{BS16}. This opens up a
possibility how microstructures -- which can be manipulated by varying the
membrane tension -- can be used to control the lateral organization of
nanodomains \cite{BS16}.

\section{Domain formation due to dynamical 
interactions with the membrane environment}
\label{sec:environment}

\begin{figure}
\centerline{
  \includegraphics[width=0.3\textwidth]{\dir/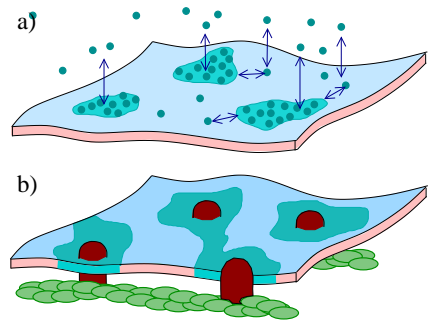}
}
\caption{
Mechanisms of nanodomain formation based on interactions with an environment.
(a) Domain stabilization due to membrane recycling. Membrane components
  (e.g., proteins or certain raft lipids such as cholesterol)
  can diffuse within the membrane, leave the membrane 
  and be reinserted (e.g., due to vesicle budding and fusion). 
  The membrane environment provides a reservoir.
  (After  \protect\cite{Foret12}).
(b) Domain stabilization due to a pinning of membrane inclusions
  to the external cytoskeleton network. (After \protect\cite{GBR13}).
}
\label{fig:fig10}       % Give a unique label
\end{figure}

So far we have considered isolated membranes, without accounting for their
interactions with the environment. Real biomembranes are embedded in a cellular
context that cannot be neglected. A number of mechanisms have been proposed
that inhibit large scale demixing in biomembranes below an equilibrium demixing
point through interactions with the cellular environment. 
They mostly belong to one of two categories:
Mechanisms related to the perpetual lipid turnover in the membrane, and
mechanisms based on the coupling of the membrane with the cytoskeleton.
Examples of these two mechanisms are sketched in Fig.\ \ref{fig:fig10}.

\subsection{Membrane recycling}
\label{sec:recycling}

%  \cite{GI99} Edidin vesicle traffic
%  \cite{TSS05} Turner lipid turnover\\
%  \cite{Foret05} Foret lipid turnover arrests phase separation (no noise)\\
%  \cite{GSR08} Gomez cholesterol turnover\\
%  \cite{DMC11} Maiti lipid turnover + proteins\\
%  \cite{Foret12} Cluster formation due to active channels (GL)
%  \cite{GKR16} lipid turnover, outflow of cholesterol, GL equation
%  \cite{SMT13} lipid turnover on two leaflets, only one phase separating\\
%
%  \cite{FSH08} Fan lipid turnover due to vesicle trafficking
%     cluster also stabilized due to immobilized proteins\\
%
%  \cite{NMW11} arrest of domain coarsening due to dynamic asymmetry:
%     bottom layer slower (diffusive), top faster (hydrodynamics), LB simulation
%
%  \cite{CC06} Chen active two-state inclusions (change shape)
%
%  \cite{GGS12} Rao active remodeling of cortical actin controls domains\\
%  \cite{RM14} Rao current opinion, underlines role of activity\\
%
%  \cite{GGS12} Rao active remodeling of cortical actin controls domains\\
%  \cite{RM14} Rao current opinion, underlines role of activity\\

We will first discuss the first category, i.e., domain formation due to
membrane recycling. In living cells, membranes constantly exchange lipids and
other membrane components with their environment, e.g.,  through vesicle
budding and fusion. The fact that vesicle traffic may explain membrane
patchyness was first pointed out in 1999 by Gheber and Edidin \cite{GI99}, and
this has triggered a series of theoretical work on this issue.

The most intensely discussed mechanism is the Foret mechanism, which describes
membrane recycling in terms of a dynamical lipid exchange with an external
reservoir. In a seminal paper of 2005 \cite{Foret05}, Foret proposed to model
membrane recycling by a dynamic Ginzburg Landau theory that accounts for
diffusive in-plane phase separation and includes an additional sink/source term
describing lipid exchange (see Eq.\ (\ref{eq:gl_dphidt}) in 
\ref{app:gl_nonequilibrium}). He studied the phase separation dynamics in this
model by computer simulations and showed that the initial domain growth is
indeed arrested due to the sink term. Estimating the final size of the domains
from the characteristic time at which the domain growth stops, he obtained an
upper limit $l^*$ which scales as $l^* \propto \tau_r^{1/3}$ with the inverse
rate $\tau_r$ of lipid exchange.

As shown in the appendix \ref{app:gl_nonequilibrium} (see also \cite{GKR16}),
the Foret model can be mapped on an established model for equilibrium
copolymeric microemulsions, the Ohta-Kawasaki model \cite{OK86}. In the
Ohta-Kawasaki model, global phase separation is suppressed by a Coulomb-type
long-range interaction term. In the Foret model, an effective long-range
interaction is mediated through the lipid exchange with an omnipresent lipid
reservoir.  The scaling of the characteristic domain size for the Ohta-Kawasaki
model is well-known \cite{OK86} and by analogy, one would expect the
characteristic domain size to scale as $l^* \propto \tau_r^{1/4}$ in the Foret
model (see \ref{app:gl_nonequilibrium}).  This prediction has not yet been
tested explicitly in the literature, but it seems to be compatible with all
published data referenced below.

Most later studies of membrane recycling effects were based on the Foret model.
Gomez \etal \cite{GSR08} applied the Foret model to cholesterol turnover and
carried out a linear stability analysis as well as further computer
simulations. According to the linear theory, the most unstable wavelength in
the initial stage of phase separation scales as $l^* \propto \tau_r^{1/2}$ with
the inverse turnover rate $\tau_r$. Motivated by this result, they fit their
simulation results for the final domain sizes $R_f$ by a linear law $R_f^{-1} =
a + b \tau_r^{-1/2}$. However, their data seem equally compatible with a fit to
$R_f = c \: \tau_r^{1/4}$.  

Das \etal \cite{DMC11} tested the Foret model against
experimental data obtained by FRET (F\"orster resonance energy transfer)
imaging of living human cancer cells. The lipid recycling times were tuned
using an ATP inhibition method. They showed that the experimental results could
be fitted quantitatively with simulation predictions of the Foret model.
Sornbundit \etal \cite{SMT13} applied the Foret model to a system of two
coupled monolayers, where phase separation occurs on one side only, and show
that this leads to domain formation on both sides.  Garcke \etal \cite{GKR16}
presented a detailed mathematical analysis of the Foret model and pointed out
the relation to the Ohta-Kawasaki functional.

Almost in parallel to Foret, in 2005, Turner \etal \cite{TSS05} developed a
closely related model which was based on a master equation for concentrations
$c_n$ of clusters containing $n$ ''monomers'' (raft proteins or other raft
components) to describe cluster growth in the presence of lipid turnover.  The
model accounted for thermodynamically driven fusion of domains and included an
additional term describing the addition or removal of clusters, which takes the
role of the source/sink term in the Foret model. Using computer simulations,
Turner \etal\ showed that this model also produces finite size domains with
characteristic domain sizes $n^*$ that increases according to $n^* \propto
1/\sqrt{j}$ with the lipid exchange rate $j$.  With the identifications $n^*
\propto {l^*}^2$ and $j = 1/\tau_r$, this result is in very good agreement with
the prediction $l^* \propto \tau_r^{1/4}$ of the Foret model .  The Turner
model was later refined by Foret \cite{Foret12}, who explicitly accounted for
the energy dissipation due to lipid turnover and identified three steady-state
regimes depending on the choice of parameters: Macroscopic phase separation,
Formation of finite domains, and a ''disordered state'' with isolated monomers.

\begin{figure*}
\centerline{
  \includegraphics[width=0.8\textwidth]{\dir/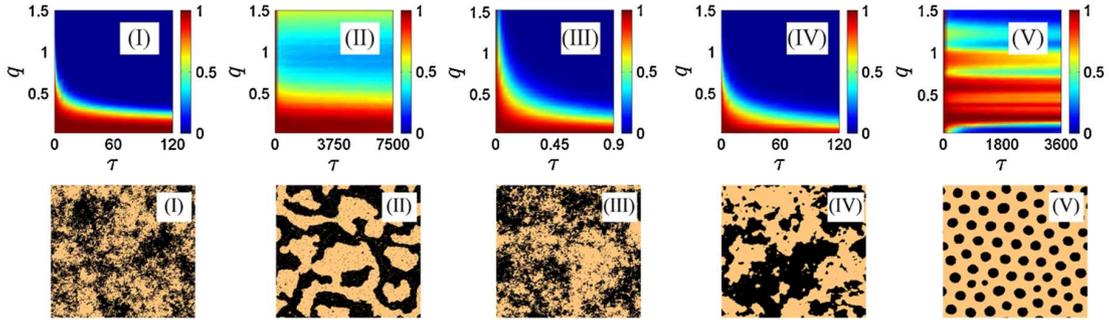}
}
\caption{
Domains emerging from different types of dynamically stabilized 
microdomains within the unified continuum treatment of Fan \etal
\protect\cite{FSH10c}.
Upper panel shows time dependent correlation function in Fourier
space, lower panel representative snapshots corresponding to
(I) Critical clusters (see Sec.\ \protect\ref{sec:critical})
(II) Random quenched pinning sites (see Sec.\ 
  \protect\ref{sec:cytoskeleton})
(III) Stochastic recycling above $T_c$, 
(IV) Stochastic recycling below $T_c$, 
(V) Foret mechanism.
See \protect\ref{app:gl_nonequilibrium} for the definition
of the underlying models.
Reprinted figure with permission from
  J. Fan \etal, 
{\em Lipid microdomains: Structural correlations,
   fluctuations, and formation mechanisms},
Phys. Rev. Lett. {\bf 104}, 118101 \protect\cite{FSH10c}.
Copyright (2010) by the American Physical Society.
}
\label{fig:fig11}       % Give a unique label
\end{figure*}

Fan \etal\ in 2008 \cite{FSH08} proposed an alternative Ginzburg-Landau model of
membrane recycling, where effect of lipid turnover is incorporated at the level
of a nonlocal active noise (see \ref{app:gl_nonequilibrium}). In this model,
the lipid redistribution is restricted to a finite spatial range.  In contrast
to the Foret model, it cannot be mapped to an equivalent equilibrium model,
hence it describes a truly nonequilibrium system. In a later publication
\cite{FSH10c}, Fan \etal\ compared the morphologies obtained with this
''stochastic recycling'' model with morphologies obtained from other models,
i.e., critical clusters, the Foret mechanism, and arrested phase separation due
to pinning sites (see next section). Examples of configurations are reproduced
in Fig.\ \ref{fig:fig11}. The Foret mechanism is the only one that produces
nearly regular arrays of clearly separated domains. In the stochastic 
recycling of Fan \etal, the clusters are much less ordered and 
less well-defined. Fan \etal \cite{FSH10c} suggested that this insight could 
be used to distinguish between different membrane recycling mechanisms
in experiments.

\subsection{Cytoskeleton coupling}
\label{sec:cytoskeleton}

%  \cite{YW07} Yethiraj disorder - interface coupling\\
%  \cite{FSH08,FSH10b} Fan lipid turnover due to vesicle trafficking
%     compartmentalization of lipids either by mesh 
%     or immobilized proteins (bulk coupling)\\
%  \cite{FSH10c} Fan et al PRL comparison of dynamical models
%     disorder - interface coupling\\
%  \cite{MPS11} Machta cytoskeleton - bulk coupling\\
%  \cite{SSK14} Laradji cytoskeleton - bulk coupling
%  \cite{LWC06} Lenne experimental reduction of bulk diffusion due
%    to cytoskeleton
%    - Vink: Bulk pinning -\\
%  \cite{FV11} Vink quenched protein obstacles universality RFIM\\
%  \cite{SV12} Vink Pink model domains in main transition on rough support\\
%  \cite{FRV12} immobilized particles \\
%  \cite{GBR13} Gomez actin network simulation\\
%  \cite{HSK14} actin meshwork organizes phase separation - experiments
% \cite{WBV12} membrane bound proteins slow down coarsening

The second important group of externally driven domain stabilization mechanisms
comprises mechanism that depend on the coupling between a plasma membrane and
the cytoskeleton \cite{KSK11}. An example is sketched in Fig.\
\ref{fig:fig10}b).  Since one important effect of the cytoplasm is to impose
disorder on the membrane by immobilizing certain membrane inclusions (e.g.,
anchor proteins) which then serve as pinning sites, most mechanisms in this
category can also be categorized as mechanisms based on quenched disorder.

Yethiraj and Weisshaar were the first to point out that global phase separation
can be suppressed and domains can be forced to break up if the membrane is
filled randomly with immobile inclusions \cite{YW07}. They demonstrated this by
Monte Carlo simulations of a simple lattice model, where randomly placed
particles were added that acted as pinning sites for domain boundaries.  We
will refer to such impurities as ''interface coupled'' pinning sites. A similar
situation was later studied by Fan et al \cite{FSH10c} using a dynamic
Ginzburg-Landau model (see \ref{app:gl_nonequilibrium}). Immobile
interface-coupled inclusions reduce the line tension, which shifts the critical
temperature to lower values \cite{YW07}. Several researchers have considered an
alternative scenario where the pinning sites have an affinity to one of the
coexisting phases \cite{FSH08, FV11, SV12, FRV12, FSH10b, WBV12}. We
will denote such impurities as  ''bulk coupled'' pinning sites, 
see Fig.\ \ref{fig:fig10}b) for an illustration.  Bulk coupled
pinning sites tend to nucleate domains of this phase, which suppresses global
phase separation.

These observations can be understood within the theory of critical phenomena:
Vink and coworkers \cite{FV11} pointed out that membranes with bulk coupled
random pinning sites can be associated with the so-called random field Ising
model (RFIM) \cite{IM75, AW89}, which is one of the archetype models for
studies of phase transitions in disordered systems.  The RFIM is derived from
the Ising model (IM), which consists of an array of interacting ''magnetic
spins'' and exhibits a continuous ordering transition characterized by the same
critical properties than the demixing transition (it belongs to the same
''universality class'').  In the RFIM, disorder is introduced by exposing the
spins to random, but fixed  ''magnetic fields'' \cite{IM75}. It has been shown
that the presence of such fields, however weak, fully suppresses the phase
transition \cite{AW89}. Vink argued that a membrane filled with randomly
distributed, bulk coupled immobile obstacles should belong to the RFIM
universality class, hence the demixing transition in such membranes should be
fully suppressed as well. He and coworkers tested this hypothesis by
simulations of various membrane models with different levels of
coarse-graining, ranging from lattice models to realistic chain models
\cite{FV11, SV12, FRV12}. They consistently found that these system no longer
phase separate in the presence of the obstacles. Phase separation was restored
if the obstacles did not favor particular phases, or if they were arranged on a
regular lattice \cite{FV11}.  

By a similar analogy, membranes with interface coupled pinning sites can be
associated with the class of random bond Ising models (RBIM), which is another
archetype for disordered systems where the disorder is introduced at the level
of the interaction between spins. In RBIM models, the strength of disorder must
exceed a critical threshold before the ordering transition is fully suppressed.
Weak disorder only moves the transition point to lower temperatures
\cite{DD83}. This is consistent with the findings of Yethiraj and Weisshaar
\cite{YW07}. Hence we conclude that, somewhat unexpectedly, bulk-coupled random
pinning sites that have a special affinity for one of the coexisting phases
suppress the demixing transition more effectively than interface-coupled random
pinning sites that attract domain boundaries.

In the studies discussed so far, the pinning sites were distributed randomly on
the membrane. Other groups have modelled the cytoskeleton network more
realistically by introducing an underlying mesh to which the pinning sites are
attached \cite{FSH10b, MPS11, HSK14}. The mesh thus organizes the pinning sites
and compartmentalizes the membrane by erecting ''fences'' of pinning sites. In
this context, an additional dynamic coupling mechanism between cytoskeleton and
membrane may also becomes important \cite{FSH10b,EPS11}: The experimentally
observed fact that the diffusivity of lipids is reduced in the vicinity of the
cytoskeleton \cite{LWC06, HSK14}.  Based on analytical arguments and
Ginzburg-Landau simulations, Fan \etal \cite{FSH10b} predicted that both this
diffusion-based mechanism and the standard bulk coupling mechanism stabilize
domain structures that are strongly correlated with the underlying cytoskeleton
structure. Machta \etal \cite{MPS11} studied the bulk coupling case by lattice
simulations and confirmed the predictions of Fan \etal \cite{FSH10b}.  They
also considered the effect of bulk coupling on lipid diffusivity and concluded
that bulk coupling to a cytoskeleton can account for diffusion confinement in
membranes close to the critical demixing point of the bare membrane, or if the
density of pinning sites is very large.  Honigmann \etal \cite{HSK14} recently
combined fluorescence correlation spectroscopy and superresolution stimulated
emission depletion (STED) imaging for an extensive experimental study of
supported bilayers coupled to an actin network. They showed that, indeed, the
actin network disrupted the formation of phase separated domains over the whole
range of considered temperatures.  However, when carrying out simulations of
the Machta model with simulation parameters matched to the experiments, they
found that the standard bulk coupling mechanism alone is not sufficient to
explain the experimental findings. The discrepancy could be resolved by
postulating an additional, curvature-based coupling between the membrane and
the actin \cite{HSK14}. 

Sikder \etal \cite{SSK14} identified yet another mechanism through which a
cytoskeletal meshwork in close proximity to a membrane can suppress phase
separation even far below the critical region: In their model, the actin
filaments do not directly influence the lipid membrane, but they locally stop
the motion of membrane-bound mobile proteins. If these proteins have an
affinity to one membrane phase, this also generates small and relatively stable
domains.
 
All these mechanisms explain domain formation by an interaction of a membrane
with an immobile cytoskeleton network below a demixing point. In living cells
at physiological temperatures,  membranes are more likely maintained
above the demixing point (see the discussion in Sec.\ \ref{sec:critical}), and
the cytoskeleton is not immobile, but subject to constant remodeling
\cite{RM14}. Recent theories of cytoskeleton-induced domain formation are
beginning to take this into account\cite{GGS12,GBR13}. Gomez \etal \cite{GBR13}
proposed a simple model where bulk-coupled pinning sites occasionally detach
from the cytoskeleton due to the repolymerization of the actin meshwork and
reattaches elsewhere. They show that in the presence of other larger mobile
bulk-coupled membrane inclusions, transient domains may form even above the
lipid demixing point. A similar mechanism of protein nanocluster formation was
studied by Gowrishankar \etal \cite{GGS12} using a more detailed and realistic
model for the dynamically evolving actin network and including hydrodynamic
interactions. 

The possible role of the cytoskeleton for domain formation has also
been highlighted by recent experimental work of Kraft \etal \cite{FLK13}
on sphingolipid domains in plasma membranes of fibroblasts. They found
the these domains (which are -- as a side note -- {\em not} enriched with 
cholesterol \cite{FKL13}) seem to depend on the coupling to the cytoskeleton
and disappear if the cytoskeleton is disrupted.
 
\subsection{Other nonequilibrium mechanisms}
\label{sec:ne_other}

Other nonequilibrium processes are conceivable that should also lead to 
domain formation in membranes. 

For example, Chen and Chen \cite{CC06} have considered the effect of membrane
inclusions (e.g., protein channels) that actively switch between two different
states. They studied a situation where domain-bound inclusions switch
stochastically between one inert state and one state where they locally deform
the membrane and attract each other. The curvature-induced and direct
interactions trigger protein aggregation, which is however arrested due to the
stochastic switching. Mathematically, this mechanism is similar to the Foret
mechanism \cite{Foret05} described in Sec.\ \ref{sec:recycling}, and the
phenomenology is also similar: One observes relatively ordered arrays of
domains whose size decreases as a function of switching rate. 

Ngamsaad \etal \cite{NMW11} suggested a rather different mechanism that can
arrest phase separation in plasma membranes at least on intermediate time
scales. They considered the phase separation on a bilayer with strongly
asymmetric demixing dynamics on each monolayer. The dynamics on one leaflet,
representing the outer leaflet of a plasma membrane, was assumed to be fast and
accelerated by hydrodynamics, whereas the dynamics on the other, inner leaflet
was taken to be slow and purely diffusive, i.e., hydrodynamic interactions were
taken to be screened due to the presence of the cytoplasm.  In this case, the
domain growth on the outer leaflet is arrested by the coupling to the smaller
domain pattern on the inner leaflet. The mechanism has similarity to the
pinning effects described in \ref{sec:cytoskeleton}.  Unless some real pinning
takes place as well, however, further coarsening takes place on large time
scales which is governed by the slow dynamics on the inner leaflet.
 
In general, the fact that phase transition kinetics can be controlled from 
the outside in membranes offers rich opportunities for generating complex
transient domain structures. In another example due to Shimobayashi \etal
\cite{SIT16}, the transition from macro- to microphase separation was studied
experimentally for ternary demixed vesicles that were exposed to a solution
containing glycolipid micelles. As the glycolipids were gradually
incorporated in the membrane, this triggered a transition to a modulated state
{\em via} an intriguing intermediate pattern of ring like structures on the
vesicle. 

\section{Conclusions and Outlook}
\label{sec:conclusions}

Summarizing, we have reviewed a large variety of mechanisms that
can contribute to heterogeneities and domain formation in multicomponent
membranes, some operating already at equilibrium and some in the nonequilibrium
context of a living organism. Given the fact that biomembranes are part of the
highly dynamic and sophisticated system of the cell, it would seem much more
surprising if they turned out to be homogeneous than if they were
heterogeneous. 

The overview shows that lipids presumably contribute substantially to such
heterogeneities, since micro- and nanostructuring seems to be a generic,
omnipresent phenomenon in multicomponent lipid membranes. Thus the question
''whether'' or ''why'' rafts exist may not be well-posed. The more relevant
question is probably: How does nature take advantage of the existence of rafts,
which seem to be imposed by the (bio)physics of membranes and hard to avoid?
Nature must find ways to {\em control} raft formation.  Since most domain
forming mechanisms are related to phase transitions, they can be manipulated by
controlling the phase behavior. Control parameters are not only provided by the
lipid composition, the temperature and the composition of the surrounding
medium, but also by mechanic factors such as the hydrostatic pressure
\cite{BS16,SS13} and the membrane tension \cite{AKZ07, NWN10, RMM11, SS13,
BS16}, and by geometric factors such as the membrane curvature (see Sec.\
\ref{sec:curved}). Furthermore, we have seen that domain forming mechanisms
interact with each other, suppress each other or promote each other. This
provides nature with a rich toolbox that can be exploited to pre-structure and
organize membranes already at the level of their physical properties, in order
to optimize them for their role in the cells.

\section*{Acknowledgments}

The contribution of our own group to this field were mostly carried out by
Sebastian Meinhardt and Leonie Brodbek, based on earlier work by Dominik
D\"uchs, Olaf Lenz, Beate West, J\"org Neder, and Stefan Dolezel, and in
collaboration with Richard Vink, Laura Toppozini, and Maikel Rheinst\"adter. We
have also benefitted from inspiring interactions with many others, in
particular Frank Brown, Markus Deserno, Gerald Feigenson, Sarah Keller, Ben
Machta, and Michael Schick. Our work was funded by the Deutsche
Forschungsgemeinschaft within the Sonderforschungsbereich SFB 613 and SFB 625.

\begin{appendix}

\section{Ginzburg-Landau theory as a theoretical framework 
  for the description of membrane domains}
\label{app:gl}

In the following, we provide a unified mathematical description of the
phenomena discussed in the main text, using the framework of the
Ginzburg-Landau theory, which is suitable to describe generic phenomena related
to phase transitions \cite{HK15}. Ginzburg-Landau approaches are commonly used
in theoretical descriptions of order/disorder phenomena in lipid membranes,
see, e.g., Ref. \cite{KA14} for a recent review. The basic idea of the
Ginzburg-Landau approach is to (i) define an ''order parameter'' that describes
the phenomenon of interest and is typically a density of an extensive quantity,
(ii) make an Ansatz for a free energy functional by expanding a local free
energy density in terms of powers of the order parameter and its spatial
derivatives, taking into account the symmetries of the system, and finally
(iii) minimize this free energy functional with respect to the order parameter
to calculate (mean-field) phase diagrams, or use it as input in a statistical
field theories that allow to assess the effect of fluctuations on the phase
behavior. We will now give a brief overview how this approach can be used to
describe domain formation in membranes. 

\subsection{Phase separation and critical fluctuations}
\label{app:gl_critical}

In the framework of the Ginzburg-Landau theory, fluid-fluid phase separation on
a planar membrane can be described by a single scalar order parameter
$\Phi({\bf r})$ describing the local composition in the two phases,
and the Ginzburg-Landau free energy functional takes the simple form
\begin{equation}
\label{eq:gl_phaseseparation}
{\cal F}[\Phi] = \int \ud^2 r \: \Big\{
\frac{t}{2} \Phi^2 + \frac{1}{4} \Phi^4 + \frac{g}{2} (\nabla \Phi)^2
+ h \Phi
\Big\}.
\end{equation}
Here the integral $\int \ud^2 r \cdot$ runs over the membrane plane, the cubic
term in the expansion has been removed by appropriately shifting the origin of
$\Phi$, $\Phi$ has been rescaled such that the fourth order term has the fixed
coefficient $1/4$, and higher order terms have been neglected. The coefficients
$t,g,h$ are phenomenological parameters, which depend on the physical control
parameters temperature, pressure, and chemical potentials.  For example, the
parameter $t$ can loosely be associated with a shifted and rescaled
temperature, the parameter $h$ controls the composition and takes the
role of a chemical potential difference, and the parameter $g$ controls the
line tension. 

Minimizing ${\cal F}$ with respect to $\Phi({\bf r})$, one finds that all
compositions $\Phi$ are possible for $t>0$, but a miscibility gap opens up for
$t < 0$, where two phases with compositions $\overline{\Phi}_{\pm} = \pm
\sqrt{-t}$ coexist with each other.  The system has a critical point at $t=0$
which belongs to the Ising universality class. Close to the critical point, the
correlation length diverges according to $\xi \propto \sqrt{g/|t|}$ in mean
field approximation. Right above the critical point, the system therefore
contains large correlated ''domains'', which have no sharp boundaries and a
broad distribution of domain sizes. Right below the critical point, interfaces
between coexisting phases become increasingly fuzzy (the interfacial width is
set by $2 \xi$) and the line tension vanishes according to $\lambda \propto \xi
t^2$. This qualitatively explains the properties of critical clusters described
in Sec.\ \ref{sec:critical}. Quantitatively, the exponents of the power laws 
deviate from the mean-field prediction due to the effect of thermal
fluctuations. In reality, the order parameter at coexistence, the correlation
length, and the line tension scale as $\overline{\Phi} \sim 
|t|^{1/8}$, $\xi \sim |t|^{-1}$, $\lambda \sim |t|^{1}$ in two dimensional
Ising-like systems. These exponents have indeed been measured within the error
in multicomponent membranes \cite{HCC08, HVK09}. We note that at the level
of these power laws, the exact interpretation of the order parameter $\Phi$ 
does not matter. Most quantities that depend linearly on local compositions
and drop to zero at the critical point will show the same scaling.

\subsection{Modulated structures and Microemulsions: Generic theory}
\label{app:gl_microemulsion}

%\cite{Brazovskii75} Brazovskii paper
%\cite{HS95} Hohenberg Swift fluctuation-driven first order transitions
%\cite{HLS75,Hornreich80} Lifshitz point
%\cite{OK86,OK90} Ohta-Kawasaki

To describe microemulsions and modulated phases within the Ginzburg-Landau
framework, the gradient contribution $(\nabla \Phi)^2$ in Eq.\
\ref{eq:gl_phaseseparation} must be generalized. A typical Ansatz includes
terms with higher order spatial derivatives of $\Phi$, which are however still
quadratic in $\Phi$ \cite{GompperSchick94}. The generic form of the free
energy functional is hence most conveniently written in Fourier representation
\begin{eqnarray}
\label{eq:gl_microemulsion}
{\cal F}[\Phi] &=& \int \ud^2 r \:  \big\{
\frac{t}{2} \Phi^2 + \frac{1}{4} \Phi^4 \big\}
\\ \nonumber &&
 +  \frac{(2 \pi)^2}{A} \sum_{\bf q}
\frac{1}{2} \tilde{G}({\bf q})  \: |\tilde{\Phi}_{\bf q}|^2,
\end{eqnarray}
where ${\bf q}$ are the Fourier wave vectors, $\tilde{\Phi}_{\bf q}$
corresponds to the Fourier transform of $\Phi({\bf r})$, $A$ is the membrane
area, and $\tilde{G}({\bf q})$ subsumes all gradient terms.  For example, Eq.\
(\ref{eq:gl_phaseseparation}) corresponds to the case 
$\tilde{G}({\bf q}) \equiv g \: q^2$. 

To investigate domain formation in the system defined by Eq.\
(\ref{eq:gl_microemulsion}), we must analyze the quantity $\Gamma({\bf q})= t +
\tilde{G}({\bf q})$, which corresponds to the inverse structure factor in a
reference homogeneous disordered (mixed) state. This homogeneous state is
unstable towards order parameter modulations if the minimum value of $\Gamma$
is negative. Let $q^*$ denote the wave vector that minimizes $\tilde{G}({\bf
q})$. At $t = - \tilde{G}({\bf q})\big|_{q^*}$, mean field theory thus predicts
a continuous phase transition from a mixed and homogeneous state, either to a
demixed state if $q^*=0$, or to a modulated state with the well-defined
characteristic wave vector $q^*$ if $q^* \ne 0$.  This latter transition
belongs to the so-called Brazovskii universality class, and it has the
remarkable feature that fluctuations not only shift the transition point to
lower $t$, but also change it into a first order transition
\cite{Brazovskii75,HS95}. Above the transition, the system is thus globally
disordered, but it is still ordered on local scales and filled with domains
with the characteristic length scale $2 \pi/q^*$. Such a local order
characterizes a microemulsion. It sets in at the so-called Lifshitz line where
$\Gamma({\bf q})$ first develops a minimum at a nonzero wave vector ${\bf q}$
\cite{GompperSchick94}.

Specific choices of $\tilde{G}({\bf q})$ have received particular interest in
the literature. The simplest and most prominent choice is $\tilde{G}({\bf q}) =
g \: q^2 + k q^4$ with $k>0$.  For $g>0$, this system exhibits a regular
Ising-type demixing transition at $t=0$, and for $g<0$, it undergoes a
Brazovskii transition to a modulated phase with a characteristic wave vector
$q^* = \sqrt{-g/2k}$ at $t = g^2/4 k$.  The Brazovskii and the Ising line meet
at $t=g=0$ in a multicritical point, the so-called Lifshitz point \cite{HLS75,
Hornreich80}, which is also the end point of a (first order) triple line
separating the two demixed phases (with homogeneous order parameter
$\overline{\Phi}_{\pm}$) and the modulated phase at low $t$.  Like the rest of
the Brazovskii line, the Lifshitz point is also unstable with respect to
thermal fluctuations. Computer simulations \cite{SMS14, SMV14} suggest that it
is shifted towards larger $g$ and turns into a tricritical point, i.e., the
demixing transition becomes first order and the demixing line meets the (first
order) Brazovskii line in a quadruple point.

The Lifshitz point scenario is relevant for the domain forming mechanisms
built on linactants (Sec.\ \ref{sec:linactants}) and on bilayer
curvature coupling (Sec.\ \ref{sec:bilayer_curvature}), see the discussion
below in \ref{app:gl_linactant} and \ref{app:gl_curvature}. Figs.\
\ref{fig:fig5} and \ref{fig:fig6} show mean field phase diagrams that are
typical for this scenario.  It is worth noting that the characteristic wave
length $2 \pi/q^*$ diverges at the Lifshitz point, hence the characteristic
domain sizes can become very large. 

Scenarios where $q^*$ always stays finite are also conceivable, but they
require more complex expressions for $\tilde{G}({\bf q})$. This is the case,
e.g., in the monolayer curvature coupling mechanism (Sec.\
\ref{sec:monolayer_curvature}, see \ref{app:gl_curvature}) and a typical phase
diagram is shown in Fig.  \ref{fig:fig8}. 

Another special choice of $\tilde{G}({\bf q})$ which has been quite popular in
theories for microphase separation in copolymer melts and turns out to be
relevant in the context of membrane recycling (see \ref{app:gl_nonequilibrium})
is $\tilde{G}({\bf q}) = a/q^2 + g \: q^2 $. This form was originally proposed
by Ohta and Kawasaki in the 90s \cite{OK86,OK90}.  Note that the contribution
$\tilde{G}_L = a/q^2$ no longer describes a short range interaction.  In real
space, it corresponds to a long-range Coulomb-type interaction term of the form
\begin{equation}
\label{eq:gl_Ohta_Kawasaki}
\frac{a}{2} \int \ud^2 r \: \ud^2 r' \: G_L({\bf r},{\bf r}')  \: 
\delta \Phi({\bf r}) \: \delta \Phi({\bf r}')
\end{equation}
in the free energy functional, where we have used the notation
$\delta \Phi({\bf r})  = \Phi({\bf r}) - \langle \Phi \rangle$
and  $G_L$ is the solution of
$\Delta \: G_L({\bf r},{\bf r}') = - 2 \pi \delta({\bf r} - {\bf r}')$ in two
dimensions.
This nonlocal term inhibits macroscopic phase separation and stabilizes
structured mesophases with a characteristic wave vector of $(a/g)^{1/4}$
\cite{OK86}. 

\subsection{Mechanisms that stabilize microemulsions}
\label{app:gl_me_stabilization}

We turn to the description of mechanisms that stabilize microemulsions, i.e.,
mechanisms that generate effective free energy functionals of the form
(\ref{eq:gl_microemulsion}).

\subsubsection{Line active molecules}
\label{app:gl_linactant}

The effect of line active additives can be described by adding an additional
density field $\rho_l$ representing the density of linactants, to Eq.\
(\ref{eq:gl_phaseseparation}).  The resulting coupled functional (at $h=0$)
reads
\begin{eqnarray}
\nonumber
{\cal F}[\Phi,\rho_l] &=& \int \ud^2 r \:  \Big\{
\frac{t}{2} \Phi^2 + \frac{1}{4} \Phi^4 + \frac{g}{2} (\nabla \Phi)^2
\\
&& \quad + \; \rho_l \ln \rho_l - \mu_{\mbox{\tiny eff}} \rho_l \Big\},
\label{eq:gl_linactant}
\end{eqnarray}
where $\mu_{\mbox{\tiny eff}} ({\bf r})$ depends on $\Phi$ and subsumes the
local interactions between the surfactants and the order parameter field.
Since the linactants are attracted to the domain boundaries, we can assume the
simple form $\mu_{\mbox{\tiny eff}} = \mu_l + \alpha g (\nabla \Phi)^2$, where
$\mu_l$ is the chemical potential of linactants, and $\alpha g$ describes the
strength of the attraction. Minimizing the free energy functional
(\ref{eq:gl_linactant}) with respect to $\rho_l$ and omitting terms of order
$(\nabla \Phi)^4$, we recover a functional of the form
(\ref{eq:gl_phaseseparation}), except that $g$ is replaced by $g_{\mbox{\tiny
eff}} = g(1-\alpha \rho_{l,\infty})$, where $\rho_{l,\infty} = \exp(\mu_l)$ is
the linactant density in the bulk, far from the domain boundaries.  

Thus the linactants effectively reduce the line tension. For sufficiently high
coupling constant $\alpha$, the line tension becomes negative at a critical
linactant density, and the system turns into a microemulsion {\em via} a
Lifshitz point.  The corresponding phase diagram (based on Eq.\
(\ref{eq:gl_linactant}) with an additional stabilizing term $\frac{1}{2}
(\Delta \Phi)^2$) is shown in Fig.\ \ref{fig:fig5}. This phase diagram was
produced with the parameters $g=1$ and $\alpha=1$ and a single-mode
approximation was used to calculate the first order boundary between the
modulated and the phase separated region \cite{BS16}. We note that adding
surfactants typically also reduces the driving force for phase separation,
hence $t$ should shift to higher values with increasing $\rho_{l,\infty}$.
In the main text, we propose to use the simple Ansatz
$t_{\mbox{\tiny eff}} = t_0 + \nu \rho_{l,\infty}$, which corresponds to the
leading order of $\rho_{l,\infty}$.

\subsubsection{Curvature induced mechanisms}
\label{app:gl_curvature}

Following our recent work \cite{BS16}, we discuss the bilayer and monolayer
curvature coupling mechanisms within a single unified framework. The starting
point is a description of the membrane in terms of two coupled elastic sheets,
representing the monolayers. This type of description was originally proposed
by Dan \etal \cite{DPS93, DBP94, ABD96} and has been used successfully by
others and us to describe membrane deformations due to inclusions and membrane
fluctuations \cite{BB06, WBS09, NWN10, NWN11, NNW12}. For nearly planar
membranes, the monolayer positions can be parametrized by functions
$z_{1,2}({\bf r})$ and the coupled elastic energy in quadratic approximation
can be written in the form
\begin{eqnarray}
\nonumber
%{\cal F}_{el}[z_1,z_2] 
\lefteqn{
{\cal F}_{el}
= \int \ud^2 r \:
\Big\{ \frac{k_A}{8 t_0^2} (z_1-z_2)^2
  + \frac{k_c}{4}((\Delta z_1)^2 + (\Delta z_2)^2)}
\\ \nonumber &&
  + k_c (c_1 \Delta z_1 - c_2 \Delta z_2)
  + \frac{\Gamma}{8} (\nabla (z_1 + z_2))^2
\\  &&
  + k_c \frac{\zeta}{2 t_0} (z_1 - z_2)(\Delta z_1 - \Delta z_2).
\Big\}
\label{eq:gl_elastic}
\end{eqnarray}
This expression contains all terms up to second order of $z_{1,2}$ and second
order derivatives that are allowed by symmetry, and the parameters have the
following interpretation: $t_0$ -  mean monolayer thickness, $k_A$ - area
compressibility, $k_c$ - bending energy, $c_i$  - spontaneous curvature on
monolayer $i$, $\Gamma$ - membrane tension
\cite{NWN10,Schmid11,Schmid12,Schmid13}, and $\zeta$ - an additional
curvature-related parameter.  The values of the parameters can be adjusted to
values known from experiment or simulations \cite{BB06, WBS09}. We assume that
every monolayer carries lipids that have a tendency to phase separate, hence
monolayers $i$ carry an order parameter $\varphi_i$ with a free energy
functional derived from Eq.\
(\ref{eq:gl_phaseseparation}),
\begin{equation}
\label{eq:gl_op}
{\cal F}_{\varphi}[\varphi_1,\varphi_2] = 
{\cal F}_0(\varphi_1) + {\cal F}_0(\varphi_2) - 
\int \ud^2 r \: \frac{s}{2} \varphi_1 \varphi_2
\end{equation}
\begin{displaymath}
\mbox{with} \quad 
{\cal F}_0(\varphi) =  \int \ud^2 r \: \big\{
  \frac{g}{4} (\nabla \varphi)^2
  + \frac{r}{4} \varphi^2 + \frac{1}{8} \varphi^4 \big\}.
\end{displaymath}
The parameter $s$ describes the coupling across the leaflets, i.e., domains
tend to be in registry for $s > 0$ and they tend to be staggered for $s < 0$.
The curvature coupling is implemented {\em via} $c_i({\bf r}) = c_0 + \hat{c}
\: \varphi_i({\bf r})$, i.e., the spontaneous curvature of a monolayer is taken
to depend on the local composition of the monolayer. The bilayer and monolayer
coupling mechanisms are then obtained by minimizing the total free energy,
${\cal F} = {\cal F}_{el} + {\cal F}_{\varphi}$ either with respect to a
fluctuating membrane height $h({\bf r}) =  \frac{1}{2}(z_1+z_2)$ or with
respect to a fluctuating membrane thickness parameter $u({\bf r}) =
\frac{1}{2}(z_1-z_2)$  while confining the other parameter to a constant.  This
yields in both cases a free energy functional of the form
(\ref{eq:gl_microemulsion}) for a single scalar order parameter which is a
combination of the $\varphi_{1,2}$.

In the Leibler-Andelman (bilayer curvature coupling) mechanism, the thickness
is kept constant and the free energy is minimized with respect to height
fluctuations. This results in a free energy functional ${\cal F}(\Psi)$ for the
local order parameter difference on the two leaflets, $\Psi = \frac{1}{2}
(\varphi_1 - \varphi_2)$, which has the form (\ref{eq:gl_microemulsion}) with
$t=r+s$ and
\begin{equation}
\label{eq:gl_bilayer}
\tilde{G}_{\Psi}({\bf q}) = g - 4 k_c \hat{c}^2 
  \frac{(\xi_{\Gamma} q)^2}{1+(\xi_\Gamma q)^2},
\end{equation}
where $\xi_{\Gamma} = \sqrt{k_c/\Gamma}$ \cite{LQG05,Schick12}. Expanding $G$
in powers of $q$, one obtains $G=g_{\mbox{\tiny eff}} + k q^2$ with
$g_{\mbox{\tiny eff}} = g - 4 k_c \hat{c}^2 \xi_{\Gamma}^2$ and $k = 4 k_c
\hat{c}^2 \xi_{\Gamma}^4$, which leads to the Lifshitz point scenario described
in \ref{app:gl_microemulsion}: If one decreases $t$, the system phase separates
for low curvature coupling $\hat{c}$ and/or large membrane tension $\Gamma$ and
forms a modulated structure at high $\hat{c}$ and/or low $\Gamma$. The two
regimes are connected by a Lifshitz point where the characteristic wave length
of domains diverges.  Fig.\ \ref{fig:fig6} shows a typical example of such a
phase diagram \cite{BS16}. Similar phase diagrams have been published by
Shlomovitz \etal \cite{SS13, SMS14}. We note that the characteristic length
scale $l^*$ also becomes large if the membrane tension $\Gamma$ is small, since
it scales with $\xi_{\Gamma}$.  (More precisely, the characteristic wave length
of the modulations is given by $q^* = \sqrt{\frac{\Gamma}{k_c}} \sqrt{2
\hat{c}/\sqrt{\Gamma g} -1}$).  With typical experimental values for the
bending rigidity in the range of $5-20 \cdot 10^{-20}$J \cite{Marsh06} and for
the membrane tension in the range of $\Gamma \sim 10^{-5}$N/m in
plasma membranes \cite{GMS12,PAF13}, one finds that $\xi_{\Gamma}$ is 
of the order $\sim 100$ nm.

In the monolayer curvature coupling mechanism, the height is kept constant and
the free energy is minimized with respect to thickness fluctuations. One
obtains a free energy functional ${\cal F}(\Phi)$ for the local average of the
order parameters on the two leaflets, $\Phi =
\frac{1}{2}(\varphi_1+\varphi_2)$, which has the form
(\ref{eq:gl_microemulsion}) with $t=r-s$ and
\begin{equation}
\label{eq:gl_monolayer}
\tilde{G}_{\Phi}({\bf q}) = g \: q^2 - 4 k_c \hat{c}^2 
  \frac{(\xi_0 q)^4}{(1- (\xi_0 q)^2)^2+\Lambda (\xi_0 q)^2}.
\end{equation}
The characteristic length scale $\xi_0 = (t_0^2 k_c/k_A)^{1/4}$ and the
dimensionless parameter $\Lambda = 2 - 4 \zeta \sqrt{k_C/k_A}$ are intrinsic
material constants that cannot directly be controlled externally. Assuming that
the bending rigidity $k_c$ is roughly proportional to $k_A t_0^2$, one can
expect $\xi_0$ to be of the order of the membrane thickness. Indeed, if one
inserts the elastic parameters of DPPC bilayers, one obtains $\xi_0 \approx 1$
nm, and the parameter $\Lambda$ in this case is $\Lambda \approx 0.7$
\cite{MVS13, BS16}.  

Analyzing the functional ${\cal F}(\Phi)$ in more detail, one finds again that
the system behaves like a microemulsion for high curvature coupling and turns
into a regular phase separating system for low curvature coupling.  However,
the transition from one regime to the other does not occur {\em via} a Lifshitz
point, and the characteristic wave vector of the modulations remains finite and
roughly constant for all values of $\hat{c}$, $q^* \sim 1/\xi_0$.  Hence the
typical size of domains, $2 \pi/q^*$, is always of the order of a few
nanometers. Fig.\ \ref{fig:fig8} shows a typical phase diagram for this
situation.

If one releases the constraints on the thickness or the height, one can study
the interplay of the two coupling mechanisms \cite{BS16}.  The phase diagrams
shown in Figs.\ \ref{fig:fig6}, \ref{fig:fig8}, \ref{fig:fig9} were all
produced without constraints. They show phase diagrams in the plane of the
temperature-like parameter $t$ and the dimensionless curvature parameter
$\tilde{C} = 2 \hat{c} \xi_0 \sqrt{k_c/g}$ or the dimensionless membrane
tension $\tilde{\Gamma} = \Gamma t_0/\sqrt{k_c k_A}$.  The parameters are
$\Lambda = 1.4, \xi_\Gamma = 10 \xi_0, s = - 1 \: g/\xi_0^2$ in Fig.\
\ref{fig:fig6}, $\Lambda = 0.7, \xi_\Gamma = 10 \xi_0, s =  1 \: g/\xi_0^2$ in
Fig.\ \ref{fig:fig8}, $\Lambda = 0.7, \xi_\Gamma = 10 \xi_0, s =  0.1 \:
g/\xi_0^2$ in  Fig.\ \ref{fig:fig9}a), and $\Gamma = 0.7, \tilde{C}=1, s =
0.1 \: g/\xi_0^2$, in Fig.\ \ref{fig:fig9}b).  The phase diagrams were
calculated using a single-mode approximation and further approximations
described in Ref.\ \cite{BS16}.

\subsection{Dynamics and nonequilibrium phenomena}
\label{app:gl_nonequilibrium}

The Ginzburg-Landau framework can also be used to study dynamical and
nonequilibrium phenomena \cite{HK15}. Here we follow Fan \etal \cite{FSH10c},
who introduced a dynamical model that describes in a unified manner equilibrium
critical phenomena, pinning effects, the Foret mechanism of domain
stabilization by lipid turnover \cite{Foret05}, and their own
 ''stochastic recycling'' mechanism \cite{FSH08}.
Fan \etal\ formulated the following dynamical equation for the evolution 
of the order parameter $\Phi({\bf r}, \tau)$ with time $\tau$:
\begin{equation}
\label{eq:gl_dphidt}
\frac{\partial \Phi}{\partial \tau}
= M \Delta \frac{\delta {\cal F}}{\delta \Phi({\bf r})}
  - \frac{1}{\tau_r} (\Phi - \overline{\Phi})
  + \eta({\bf r},\tau).
\end{equation}
The first term describes the diffusion of lipids with mobility $M$ in the free
energy landscape given by a Ginzburg-Landau functional, the second term
is a source/sink term that accounts for a constant exchange of
lipids with an external lipid reservoir (the Foret mechanism \cite{Foret05})
with an exchange rate $\tau_r$, and the last term subsumes thermal and active
noise. The Ginzburg-Landau functional proposed by Fan \etal\ is derived from
(\ref{eq:gl_phaseseparation}) except that it accounts for the possible presence
of immobilized inclusions $i$ acting as pinning sites for domain boundaries.
It has the form
\begin{equation}
\label{eq:gl_pinning}
{\cal F}[\Phi] = \int \ud^2 r  \big\{
\frac{t}{2} \Phi^2 + \frac{1}{4} \Phi^4 + 
\frac{g}{2} (1-\alpha \rho_p({\bf r})) (\nabla \Phi)^2
\big\},
\end{equation}
\begin{displaymath}
\mbox{where} \quad
\rho_p({\bf r}) = \frac{\pi}{\sigma^2} \sum_{i=1}^N 
\exp(- \frac{|{\bf r} - {\bf r}_i|^2}{2 \sigma^2})
\end{displaymath}
is the smeared density of pinning sites and $\alpha$ the coupling parameter 
(see also in Eq.\ (\ref{eq:gl_linactant})). Finally, the noise
is Gaussian distributed with mean $\langle \eta \rangle = 0$ and
correlator
\begin{equation}
\label{eq:gl_noise}
\langle \eta ({\bf r}, \tau) \: \eta ({\bf r}', \tau') \rangle
= - \frac{H^2}{2 \pi} \Delta K_0(\frac{|{\bf r} - {\bf r}'|}{l}) \: 
\delta (\tau- \tau'),
\end{equation}
where $K_0 (x)$ denotes the zeroth order modified Bessel function of the second
kind. This noise term accounts for ''stochastic recycling'' and is constructed
such that it conserves the order parameter on large scales, but breaks the
continuity equation on length scales smaller than $l$. Hence, it provides an
alternative description of lipid turnover, since lipids are allowed
to enter and leave the membrane on local scales while the average global
composition remains constant. In the limit $l \to 0$, the noise term reproduces
spatially uncorrelated thermal noise.  If $l$ is chosen much larger than
the equilibrium correlation length $\xi$, the noise is incompatible with
the fluctuation-dissipation theorem and this creates a true nonequilibrium 
situation.

Since the model (\ref{eq:gl_dphidt}) contains elements of several domain
forming mechanisms, they can be compared by choosing the parameters
appropriately. For example, if one chooses 
(I) $\alpha = 0, l < \xi, \tau_r^{-1}=0$, one obtains a dynamical description
for a phase separating system at equilibrium, and can study critical
clusters in the limit $t \to 0$. Starting from this system as the reference
system, one can study (II) domain pinning due to  immobilized pinning sites
by choosing $\alpha > 0$ at $t < 0$, (III, IV) stochastic recycling by choosing
$l \gg \xi$ (III) at $t> 0$ and (IV) $t < 0 $, and (V) the Foret mechanism of
lipid turnover by choosing $\tau_r^{-1} \neq 0$ at $t < 0$. Examples
of resulting morphologies are shown in Fig.\ \ref{fig:fig11} (taken from
\cite{FSH10c}). Remarkably, most mechanisms produce rather disordered,
fractal domains, except for the Foret mechanism, which generates a large
well ordered spatially periodic array of domains. This finding may seem
surprising at first, but it can be rationalized if one realizes that the 
Foret mechanism can be mapped onto the Ohta-Kawasaki model described
in Sec.\ \ref{app:gl_microemulsion} (Eq.\ (\ref{eq:gl_Ohta_Kawasaki})).
Indeed, the lipid reservoir term in Eq.\ (\ref{eq:gl_dphidt}) can 
easily be reproduced by adding a contribution of the form 
(\ref{eq:gl_Ohta_Kawasaki}) in the free energy functional
(\ref{eq:gl_pinning}) with prefactor $a = 1/M\pi \tau_r$. 
Thus the Foret mechanism effectively produces a microemulsion with a
characteristic wave vector that scales as $q^* \sim \tau_r^{-1/4}$ with
the rate $\tau_r$ of lipid exchange.

The dynamical model (\ref{eq:gl_dphidt}) can be extended to describe other and
more complex situations. For example, the free energy functional in
(\ref{eq:gl_dphidt}) can be chosen at will and supplemented with all the terms
that were discussed in the previous section. Eq.\ (\ref{eq:gl_pinning})
describes a situation where pinning sites couple to domain boundaries,
but the pinning term can easily be designed such that it describes 
pinning sites that attract one phase -- the coupling term then takes the 
simple form $\int \ud^ 2 r \: \alpha \rho_p \Phi$ \cite{GBR13}. A realistic
model for membrane dynamics should also include hydrodynamic interactions.
This can also be done, e.g., following the theoretical approaches that 
are reviewed in Ref.\ \cite{KA14}.

\end{appendix}

\section*{References}

\bibliography{membrane}

\end{document}